\documentclass{aa}  

\usepackage{graphicx}
\usepackage{float}
\usepackage{natbib}
\usepackage{booktabs}
\bibpunct{(}{)}{;}{a}{}{,}
\usepackage{txfonts}
\begin{document}

   \title{Methodological refinement of the\\ submillimeter galaxy magnification bias \\I. Cosmological analysis with a single redshift bin}
   \titlerunning{Methodological refinement of the submillimeter galaxy magnification bias. Paper I.}
   \authorrunning{Cueli M. M. et al.}

   \author{Cueli M. M.\inst{1,2}, Gonz{\'a}lez-Nuevo J.\inst{3,4}, Bonavera L.\inst{3,4}, Lapi A.\inst{1,2,5,6}, Crespo D.\inst{3,4}, Casas J. M.\inst{3,4}}

  \institute{$^1$SISSA, Via Bonomea 265, 34136 Trieste, Italy\\
    $^2$IFPU - Institute for fundamental physics of the Universe, Via Beirut 2, 34014 Trieste, Italy\\
    $^3$Departamento de Fisica, Universidad de Oviedo, C. Federico Garcia Lorca 18, 33007 Oviedo, Spain\\
             $^4$Instituto Universitario de Ciencias y Tecnologías Espaciales de Asturias (ICTEA), C. Independencia 13, 33004 Oviedo, Spain\\         
$^5$IRA-INAF, Via Gobetti 101, 40129 Bologna, Italy\\
$^6$INFN-Sezione di Trieste, via Valerio 2, 34127 Trieste,  Italy}

   \date{}

 
  \abstract
   {}
   {The main goal of this work is to test the results of a methodological improvement in the measurement of the magnification bias signal on a sample of submillimeter galaxies. In particular,  we investigate the constraining power of cosmological parameters within  the $\Lambda$CDM model. We also discuss  important points that can affect the results.}
   {We measured the angular cross-correlation function between a sample of foreground GAMA II galaxies in a single wide spectroscopic redshift bin of $0.2<z<0.8$ and a sample of background submillimeter galaxies from Herschel-ATLAS. We focused on the photometric redshift range of $1.2<z<4.0,$  with an improved methodological framework. Interpreting the weak lensing signal within the halo model formalism and performing a Markov chain Monte Carlo (MCMC) algorithm, we obtained the posterior distribution of both the halo occupation distribution and cosmological parameters within a flat $\Lambda$CDM model. Our analysis was also performed with  additional galaxy clustering information via a foreground angular auto-correlation function.}
   {We observed an overall remarkable improvement in terms of uncertainties in both the halo occupation distribution and cosmological parameters with respect to previous results. A priori knowledge about $\beta$, the logarithmic slope of the background integral number counts, is found to be paramount to derive constraints on $\sigma_8$ when using the cross-correlation data alone. Assuming a physically motivated prior distribution for $\beta$, we obtain mean values of $\Omega_m=0.23^{+0.03}_{-0.06}$ and $\sigma_8=0.79^{+0.10}_{-0.10}$ and an unconstrained distribution for the Hubble constant. These results are likely to suffer from sampling variance, since one of the fields, G15, appears to have an anomalous behavior with a systematically higher cross-correlation. We find that removing it from the sample yields mean values of $\Omega_m=0.27^{+0.02}_{-0.04}$ and $\sigma_8=0.72^{+0.04}_{-0.04}$ and, for the first time, a (loose) restriction of the Hubble constant is obtained via this observable: $h=0.79^{+0.13}_{-0.14}$. The addition of the angular auto-correlation of the foreground sample in a joint analysis tightens the constraints, but also reveals a discrepancy between both observables that might be an aggravated consequence of sampling variance or due to the presence of unmodeled aspects on small and intermediate scales.}
   {}


   \maketitle
%

\section{Introduction}

The submillimeter galaxy magnification bias was recently proposed as a novel approach to constrain cosmology via a weak lensing-induced cross-correlation between a foreground galaxy sample and a background set of submillimeter galaxies \citep{BON20,GON21,BON21}. Indeed, the phenomenon of magnification bias \citep[see][and references therein]{BARTELMANN01} can boost the flux of background sources, while increasing the solid angle they subtend. However, imposing a flux threshold effectively creates a mismatch between the two effects, which can result in an excess of background sources around those in the foreground with respect to the absence of lensing. Although it has traditionally been deemed inferior to shear analyses for the probing of the galaxy-matter cross-correlation, the realization that submillimeter galaxies provide an optimal background sample for magnification bias studies \citep[as shown by the very significant early detections of this cross-correlations in][]{WANG11,GON14} has turned this observable into an emerging independent cosmological probe. 

The reason behind their relevance for these studies lies in the fact that submillimeter galaxies are typically located at high ($z\gtrsim 1-1.5$) redshifts \citep{CHAPMAN04,CHAPMAN05,AMBLARD10,LAPI11,GON12,PEARSON13} are faint in the optical band due to thermal emission from dust and, most importantly, have a steep luminosity function \citep{GRANATO04,LAPI06,EALES10,LAPI11}. The strength of the magnification bias effect depends strongly on this last feature; indeed, the steeper the number counts, the larger the number of faint sources that may go over the detection threshold and the more likely it is that the dilution effect of magnification bias is overcome by the flux boosting. 

The current concordance cosmological model has been shown to successfully reproduce a large number of cosmological observations, namely, the cosmic microwave background (CMB) temperature and polarization spectra \citep{PLANCK14,PLANCK16,PLANCK20}, baryon acoustic oscillation measurements \citep{EISENSTEIN05,BEUTLER11,BAUTISTA21}, the primordial abundance of light nuclei \citep{CYBURT16,FIELDS20}, or the magnitude-redshift relation from type Ia supernovae \citep{PERLMUTTER99,BROUT22,SCOLNIC22,RIESS22}, among many others. However, in this day and age, the necessity for additional independent cosmological probes is unquestionable. Indeed, regardless of its countless successes, the standard cosmological model is not short of both theoretical and observational challenges \citep{BULL16,DIVALENTINO22,ABDALLA22,PERIVOLAROPOULOS22}. Regarding the latter, a special mention should be made to the ubiquitous $\gtrsim 4\sigma$ tension between local measurements of the Hubble constant derived via a distance ladder approach \citep{RIESS22} and the corresponding values from CMB anisotropy measurements \citep{PLANCK20}. Additional inconsistencies that are worth mentioning are related, for instance, to differences in the value of the structure growth parameter, $S_8\equiv \sqrt{\Omega_m/0.3}\sigma_8$, between "direct" approaches and the CMB power spectra \citep{SECCO22} or to the presence of anomalies in the anisotropies of the CMB \citep{PLANCK20b}.

Along this line, the submillimeter galaxy magnification bias has been put forward as a novel and independent cosmological probe that does not seem to suffer from the typical $\sigma_8-\Omega_m$ degeneracy found in other lensing observables. In particular, \citet{GON21} performed a first analysis and correction of the large-scale biases that could contaminate the signal and, as a consequence, the cosmological constraints. Moreover, \citet{BON21} divided up the foreground sample into four redshift bins to perform a tomographic analysis, which opened up the possibility of studying the time evolution of the dark energy equation of state. Although the Hubble constant could not be constrained, they obtained mean values of $\Omega_m=0.33^{+0.08}_{-0.16}$ and $\sigma_8=0.87^{+0.13}_{-0.12}$ at 68\% credibility within the $\Lambda$CDM model.

The present work, which is intended to be released along another companion paper, aims to build upon the aforementioned results, along with a refinement in terms of the methodology. Here, we address the cosmological and halo occupation distribution (HOD) constraints that can be derived using a single broad foreground redshift bin and updated cross-correlation data. The computation of the theoretical model for the signal is also revisited with respect to \citet{BON20} and \citet{GON21}, assessing the importance of the value of the logarithmic slope of the background galaxy number counts and of a numerical correction regarding the computation of the non-linear power spectra. In \citet{BON23}, to be referred to as Paper II, the analysis is extended to a tomographic setup, where the foreground sample is split into different redshift bins. The dependence on the number of redshift bins and their range is discussed along with the possible improvements with respect to the use of a single broad bin.

The study carried out in this work uses the measurement of the angular cross-correlation function between a sample of background submillimeter galaxies from H-ATLAS \citep{PILBRATT10,EALES10} and a sample of foreground galaxies from GAMA II \citep{DRIVER11,BALDRY10,BALDRY14,LISKE15}. Assuming a flat $\Lambda$CDM cosmology, we performed a Markov chain Monte Carlo (MCMC) analysis to derive the posterior probability distribution of both HOD and cosmological parameters. Additionally, we include the information about the clustering of the foreground sample and discuss the importance of the steepness of the submillimeter galaxy number counts.

This paper has been structured as follows. Section 2 lays out the theoretical background of this work. We discuss the physical origin of the weak lensing-induced foreground-background cross-correlation and how it is computed within the halo model formalism. The methodology followed in the paper is described in Section 3: the galaxy samples are detailed, along with the estimation procedure for both the angular cross- and auto-correlation functions and the MCMC setup to sample the posterior probability distribution of the parameters involved in each of the cases we studied. Section 4 presents the results we obtained and Section 5 summarizes our conclusions.

\section{Theoretical framework}

\subsection{Galaxy clustering and the cross-correlation induced by magnification bias}

As noted in the introduction, the weak lensing-induced foreground-background number cross-correlation is the main observable of this paper. However, a joint analysis together with the clustering of the foreground galaxy sample was also carried out in addition to study the potential tightening of the parameter constraints. Therefore, we proceed to describe the theoretical modeling of both observables below.

Under the well-known \citet{LIMBER53} and flat-sky approximations, the foreground angular auto-correlation function is given by
\begin{equation}
    w_{\text{ff}}(\theta)=\int_0^{\infty}\!\!\!dz\,\frac{H(z)}{c}\,\bigg[\frac{dN_{\text{f}}/dz}{\chi(z)}\bigg]^2\int_0^{\infty}\frac{dl}{2\pi}\,l\,P_{\text{gg}}(l/\chi(z),z)\,J_0(l\theta),\label{wff}
\end{equation}

where $P_{\text{gg}}$ is the galaxy power spectrum, $H(z)$ is the Hubble parameter at redshift $z$, $\chi(z)$ is the comoving distance at redshift $z$, $dN_{\text{f}}/dz$ is the normalized foreground source distribution, and $J_0$ is the zeroth-order Bessel function of the first kind. 

Moreover, the phenomenon of magnification bias, central to this work, probes the galaxy-mass correlation via the foreground-background angular cross-correlation function. Indeed, in the presence of lensing, the phenomenon of magnification bias produces fluctuations in the number density of background sources that, in the weak-lensing regime, can be expressed as \citep{BARTELMANN01}:
\begin{equation*}
    \delta n_{\text{b}}^{\mu}(\theta)\approx 2(\beta-1)\,\kappa(\theta),
\end{equation*}
where $\beta$ is the logarithmic slope of the unlensed background number counts\footnote{The intrinsic integral number counts of the background sources are assumed to be described by a power law in a neighborhood of the detection limit: $n_{\text{b}}(>S)=A\,S^{-\beta}$.} and $\kappa$ is the effective convergence field. Since the foreground sources trace the underlying matter field, their fluctuations are due to pure clustering, that is, 
\begin{equation*}
    \delta n_{\text{f}}^c(\varphi)=\int_0^{\infty}dz\,\frac{dN_{\text{f}}}{dz}\,\delta_{\text{g}}(\varphi,z).
\end{equation*}
Therefore, for two galaxy samples with nonoverlapping redshift distributions, the only non-negligible contribution to the foreground-background angular cross-correlation comes from the two above terms: $w_{\text{fb}}(\theta)\equiv \langle \delta n_{\text{f}}^c(\varphi)\,\delta n_{\text{b}}^{\mu}(\varphi+\theta)\rangle_\varphi$. Once again, using the Limber and flat-sky approximations, this can be expressed as \citep{cooray02}:

\begin{align}   
w_{\text{fb}}(\theta)&=2(\beta-1)\int_0^{\infty}\frac{dz}{\chi^2(z)}\frac{dN_{\text{f}}}{dz}W^{\text{lens}}(z)\,\cdot\nonumber\\
&\cdot\int_0^{\infty}dl\frac{l}{2\pi}P_{\text{g-m}}(l/\chi(z),z)J_0(l\theta)\label{wfb},
\end{align}

where

\begin{equation*}    
W^{\text{lens}}(z)\equiv\frac{3}{2}\frac{1}{c^2}\bigg[\frac{H(z)}{1+z}\bigg]^2\int_z^{\infty}dz'\frac{\chi(z)\chi(z'-z)}{\chi(z')}\frac{dN_{\text{b}}}{dz'}.
\end{equation*}

Here, $P_{\text{g-m}}$ is the galaxy-matter cross-power spectrum and $dN_{\text{b}}/dz$ is the normalized background source distribution.

\subsection{Modeling of the non-linear power spectra}

According to the halo model of structure formation, the non-linear galaxy, matter and galaxy-matter power spectra can be computed analytically if the following ingredients are provided: the halo mass function, $n(M,z)$, the deterministic linear halo bias, $b_1(M,z)$, the halo density profile, $\rho(r)$, the linear matter power spectrum, $P^{\text{lin}}(k,z),$ and the HOD. For instance, the galaxy power spectrum can be expressed as:

\begin{equation*}
    P_{\text{gg}}(k,z)=P_{\text{gg}}^{\text{1h}}(k,z)+P_{\text{gg}}^{\text{2h}}(k,z),
\end{equation*}

where $P_{\text{gg}}^{\text{1h}}$ and $P_{\text{gg}}^{\text{2h}}$ are known as the one-halo and two-halo terms and account for galaxy correlations within single halos and among different halos, respectively. They can be written as

\begin{align*}
    P_{\text{gg}}^{\text{1h}}(k,z)&=\int \,dM\,n(M,z)\,\frac{\langle N_c \rangle_M\langle N_s\rangle_M}{\bar{n}_g(z)^2}|u(k|M,z)|+\\
    &+\int\,dM\,n(M,z)\,\frac{\langle N_s(N_s-1)\rangle_M}{\bar{n}_g(z)^2}|u(k|M,z)|^2
\end{align*}

and

\begin{align*}
    P_{\text{gg}}^{\text{2h}}(k,z)=P^{\text{lin}}(k,z)\bigg[\int dM\,n(M,z)\,b_1(M,z)\frac{\langle N\rangle_M}{\bar{n}_g(z)}|u(k|M,z)|\bigg]^2,
\end{align*}

where $\bar{n}_g(z)$ is the mean number density of galaxies at redshift $z$, $u(k|M,z)$ is the normalized Fourier transform of the density profile of a typical halo of mass $M$ at redshift $z$ and $\langle N \rangle_M$ is the first moment of the HOD, that is, the mean number of galaxies in a halo of mass $M$, which has been split into the contributions of central and satellite galaxies, expressed as $\langle N_c \rangle_M$ and $\langle N_s \rangle_M$, respectively. 

Regarding the galaxy-matter cross-power spectrum, the halo model prescription is expressed as:\ 

\begin{equation*}
    P_{\text{g-m}}(k,z)=P_{\text{g-m}}^{\text{1h}}(k,z)+P_{\text{g-m}}^{\text{2h}}(k,z),
\end{equation*}
where
\begin{align*}
    P_{\text{g-m}}^{\text{1h}}(k,z)&=\int_0^{\infty} dM\,M\frac{n(M,z)}{\bar{\rho}_0}\frac{\langle N_{c}\rangle_M}{\bar{n}_g(z)}|u(k|M,z)|+\\
    &+\int_0^{\infty} dM\,M\frac{n(M,z)}{\bar{\rho}_0}\frac{\langle N_{s}\rangle_M}{\bar{n}_g(z)}|u(k|M,z)|^2
\end{align*}
and
\begin{align*}
    P_{\text{g-m}}^{\text{2h}}(k,z)&=P(k,z)\Bigg[\int_0^{\infty}dM\,M\frac{n(M,z)}{\bar{\rho}_0}b_1(M,z)u(k|M,z )\Bigg]\,\cdot\nonumber\\
    &\cdot\Bigg[\int_0^{\infty}dM\frac{n(M,z)}{\bar{n}_g(z)}b_1(M,z)\,\Big(\langle N_c\rangle_M + \langle N_s \rangle_M \,u(k|M,z)\Big)\Bigg].
\end{align*}

Although a detailed description about the ingredients of the model is given in Appendix A, it suffices to say here that we have chosen the Sheth \& Tormen model for the halo mass function and the corresponding linear halo bias derived via the peak-background split \citep{ST99}, the NFW model for the halo density profile \citep{NAVARRO97}, the usual power-law primordial power spectrum, and the three-parameter HOD model of \cite{ZEHAVI05}. For this, we have:
\begin{equation}
    \langle N \rangle_M=\langle N_c \rangle_M+\langle N_s \rangle_M=\bigg[1+\bigg(\frac{M}{M_1}\bigg)^{\alpha}\bigg]\,\Theta(M-M_{\text{min}}).\label{HOD}
\end{equation}

Therefore, within the halo model prescription, the galaxy and galaxy-matter power spectra  depend both on cosmology and the parameters of the HOD. The corresponding angular auto- and cross-correlation functions inherit this dependence  via Eqs. \eqref{wff} and \eqref{wfb}, where an additional cosmological influence is present, as well as the additional parameter $\beta$ in the case of the latter, as we discuss in Section \ref{sec:beta-section}.

These two last comments should be made before proceeding regarding the computation of the model with respect to previous works. Firstly, given the large number of integrals that have to be carried out for each angular value, we have resorted to a mean-redshift approximation in which the outermost integrals in Eqs. \eqref{wff} and \eqref{wfb} are not computed directly for all the redshift distribution, but through evaluation at the mean redshift of the sample due to the reduction of computational time by a factor of 10, so that:\ 
\begin{equation*}
    w_{\text{fb}}(\theta)\approx 2(\beta-1)\frac{W^{\text{lens}}(\bar{z})}{\chi^2(\bar{z})}\int_0^{\infty}dl\frac{l}{2\pi}P_{\text{g-dm}}(l/\chi(\bar{z}),\bar{z})J_0(l\theta).
\end{equation*}
The validity of this approximation ought to be proven in the first MCMC run, so that all subsequent analyses can be safely computed with it, which speeds up the computations dramatically.

Secondly, a crucial point should be raised regarding the computation of the two-halo term of the galaxy-matter cross-power spectrum within the halo model. As discussed by \citet{MEAD20} and \citet{MEAD21}, the evaluation of the corresponding integral poses a numerical problem, since typical halo mass functions assign a large fraction of mass to low-mass halos; this causes convergence of the integral to be extremely slow and to bias the large-scale behavior, which no longer reflects the typical linear regime. Although a more detailed analysis is made in Appendix B, where the correction procedure is explained, we stress here that overlooking this issue can induce a very strong bias on the cosmological results from a halo modeling of the signal. We have implemented the above correction for all cases in this paper.

\section{Data and methodology}

\subsection{The foreground and background galaxy samples}

The foreground and background galaxy samples have been extracted from the GAMA II \citep{DRIVER11,BALDRY10,BALDRY14,LISKE15} and H-ATLAS \citep{PILBRATT10,EALES10} surveys, respectively. Their common area covered three regions on the celestial equator at 9, 12 and 14.5 h (named G09, G12 and G15) and part of the south galactic pole (SGP) region, which amounts to a total of $\sim$ 207 deg$^2$. These are the same samples used in \citet{GON17}, \citet{BON20} and \citet{GON21}, where a more detailed discussion of the selection procedure can be found. 

\begin{figure}[h]
    \centering
    \includegraphics[width=\columnwidth]{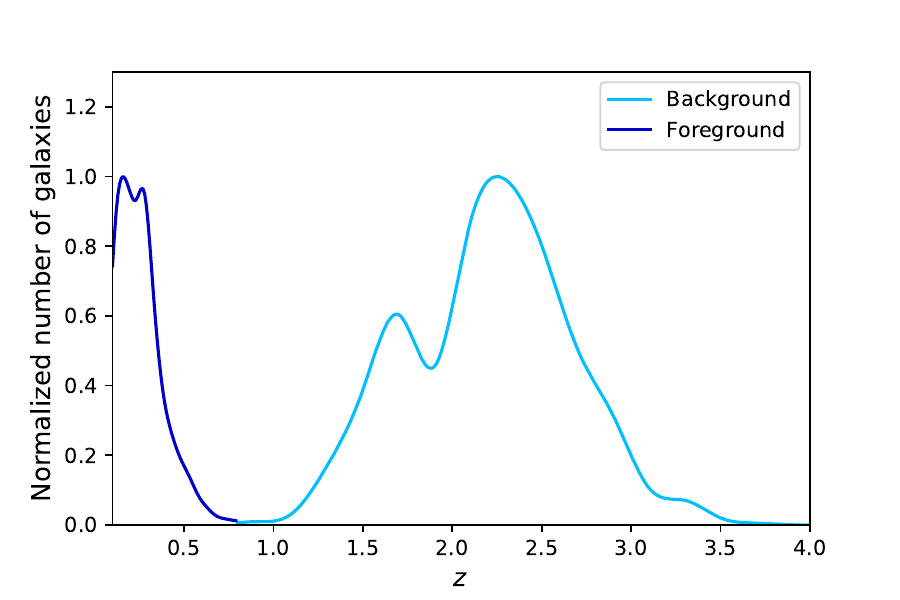}
    \caption{Normalized redshift distribution of the background (light blue) and foreground (dark blue) samples of galaxies.}
    \label{fig:dN/dz}
\end{figure}

In essence, the foreground sample is made up of GAMA II sources in the common region with H-ATLAS with spectroscopic redshifts in the range $0.2<z<0.8$, resulting in $\sim$ 130000 galaxies with a median redshift of 0.28 and surveyed down to a magnitude of $\sim$ 19.8 in the $r$ band. Figure \ref{fig:dN/dz} (in dark blue) depicts the associated redshift distribution. 

The background sample is made up of $\sim$ 37000 H-ATLAS sources in the common area, obtained via a photometric redshift selection of $1.2<z<4.0$ to ensure that there is no overlap with the foreground galaxies. The redshift estimation procedure, which is thoroughly described in \cite{GON17} and \cite{BON19}, consists of a $\chi^2$ fit of the photometry to the spectral energy distribution of SMM J2135-0102 \cite["the Cosmic Eyelash;"][]{IVISON10,SWINBANK10}, a gravitationally-lensed submillimeter galaxy at $z=2.3$ that was shown to provide the best overall fit to H-ATLAS data \citep{LAPI11,GON12,IVISON16}. The redshift distribution of the background sample, taking into account the effect of random errors, is depicted in light blue in Figure \ref{fig:dN/dz}.

\begin{figure*}[ht]
\includegraphics[width=\textwidth]{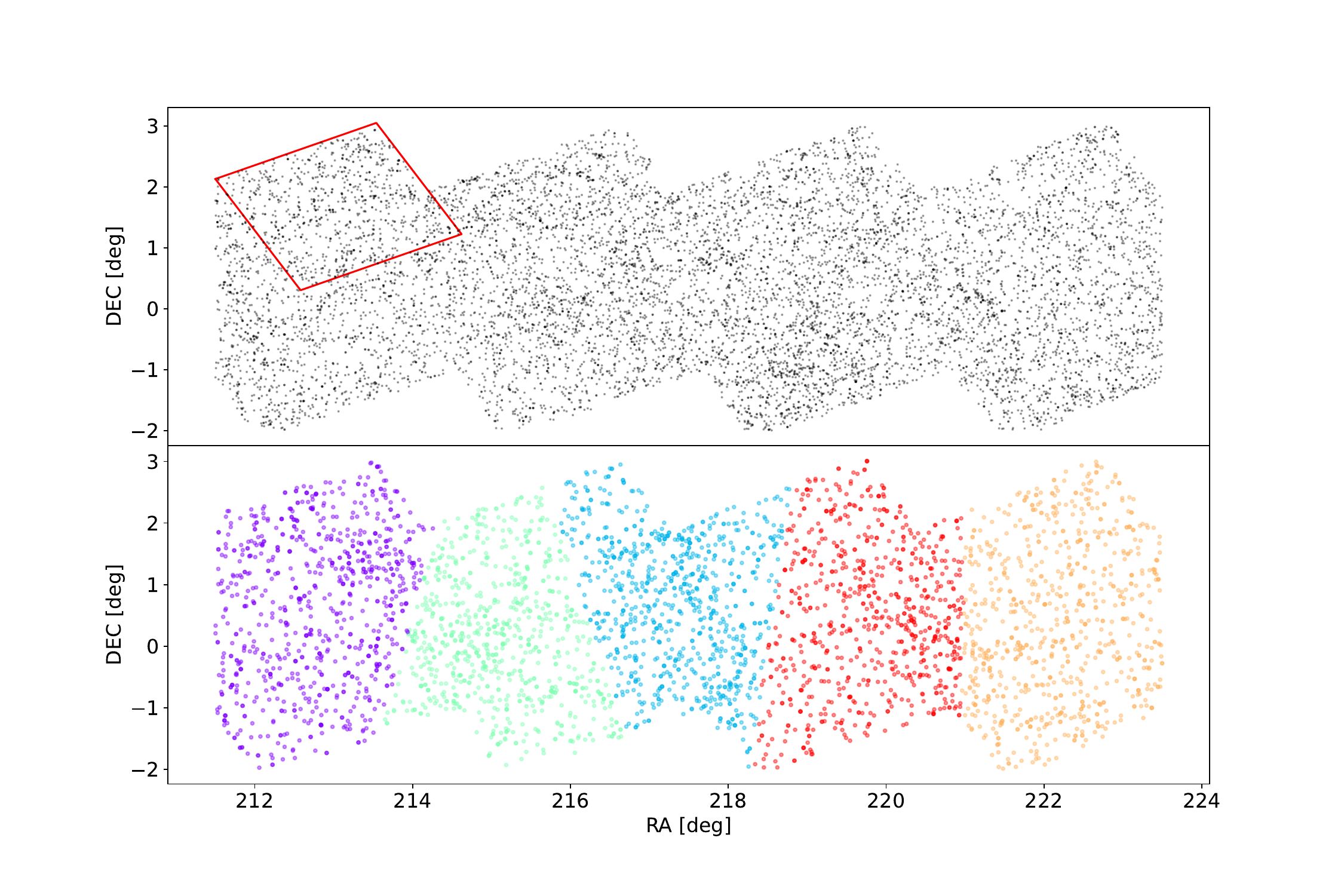}
 \caption{Angular distribution of foreground (top panel) and background (bottom panel) sample galaxies in the G15 field. In the bottom panel, different colours indicate the definition of the patches used for the Bootstrap error estimation of the angular cross- and auto-correlations. The red square in the top panel indicates the typical shape and size of a minitile, used in previous works to divide the sample into subregions (see text for more details). The number density has been artificially reduced in both panels for loading and visualization purposes.}
 \label{fig:patchesG15}
\end{figure*}

\subsection{Measurements and methodological aspects}

The scanning strategy employed by the H-ATLAS survey resulted in slightly overlapping rhomboidal shapes (or "tiles") in most fields, each of them covering about 16 deg$^2$. An average using this natural division, along with a subdivision into "minitiles" (i.e., one fourth of a tile) was analyzed in the detailed analysis of \citet{GON21}. Indeed, a cross-correlation measurement averaged over all minitiles was concluded to be the most robust method, washing out large-scale inhomogeneities and needing only the so-called integral constraint (IC) correction. The use of minitiles was, in fact, the default strategy in previous related works \citep{BON20,BON21,CUE21,CUE22}. However, as detailed in Appendix C, this approach could bias the data and, ultimately, the cosmological parameter constraints.

Therefore, and as customary in galaxy clustering studies, we now explore the possibility of performing a sole measurement of the cross-correlation function using all the available area, a methodology that should be free of IC biases given the scales probed. Although the minitiles strategy is no longer used for the measurement itself, an analogous subdivision of the whole area into minimal subregions is still necessary to assign meaningful uncertainties via internal covariance estimation. To define the subregions, we drew inspiration from TreeCorr \citep{JARVIS15}, a popular software package for computing two-point correlation functions. TreeCorr uses a k-means clustering algorithm to partition the data into subregions known as "patches," which are similar to our minitiles. Specifically, we adopted the k-means algorithm provided by the SciPy library, which aims to minimize the sum of the squared distances between data points and their assigned patch centroid. We determined the number of patches by imposing a minimum area for each of them and introduced an additional step by repeating the procedure ten times with different random initial centroids and selecting the case yielding the most consistent number of data points across patches. Figure \ref{fig:patchesG15} shows the distribution of foreground (top panel) and background (bottom panel) sample galaxies for the G15 field. A choice of (approximately equal-area) patches for the computation of the covariance matrix (as explained below) is also depicted.

\begin{figure}[h]
    \centering
    \includegraphics[width=0.48\textwidth]{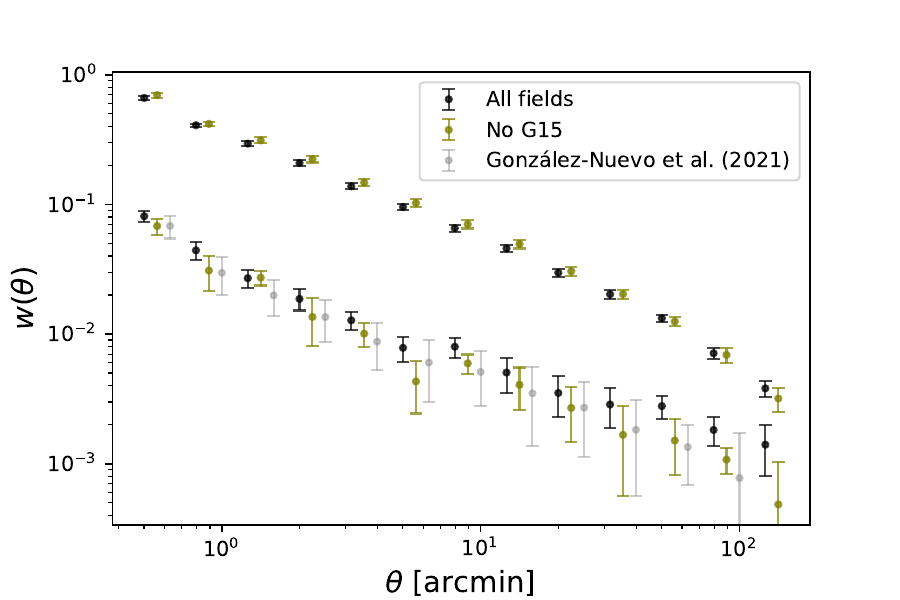}
    \caption{Measurements of the foreground auto-correlation function and the foreground-background cross-correlation function (in black) compared to the cross-correlation function excluding the G15 region (in olive green). The cross-correlation data from \cite{GON21} are shown in gray.}
    \label{xc_ac_data}
\end{figure}

The angular auto-correlation function of the foreground sample is measured over the available area through the standard \cite{LANDY93} estimator:

\begin{equation*}
    \hat{w}_{\text{auto}}(\theta)=\frac{\text{D}_{\text{f}} \text{D}_{\text{f}}(\theta)-2\text{D}_{\text{f}}\text{R}_{\text{f}}(\theta)+\text{R}_{\text{f}}\text{R}_{\text{f}}(\theta)}{\text{R}_{\text{f}}\text{R}_{\text{f}}(\theta)},
\end{equation*}
where $\text{D}_{\text{f}}\text{D}_{\text{f}}(\theta)$, $\text{D}_{\text{f}}\text{R}_{\text{f}}(\theta)$, and $\text{R}_{\text{f}}\text{R}_{\text{f}}(\theta)$ denote the normalized foreground-foreground, foreground-random and random-random pair counts at an angular separation of $\theta$, computed in practice using equally spaced logarithmic bins. The random catalog is generated from mock random positions for ten times the number of foreground sources. The measurements are shown in black in the upper part of Figure \ref{xc_ac_data}.

In turn, the foreground-background angular cross-correlation is computed  over the available area via the natural modification of the above estimator \citep{herranz01}:
\begin{equation*}
    \hat{w}_{\text{cross}}(\theta)=\frac{\text{D}_{\text{f}} \text{D}_{\text{b}}(\theta)-\text{D}_{\text{f}}\text{R}_{\text{b}}(\theta)-\text{D}_{\text{b}}
    \text{R}_{\text{f}}(\theta)+\text{R}_{\text{f}}\text{R}_{\text{b}}(\theta)}{\text{R}_{\text{f}}\text{R}_{\text{b}}(\theta)},
\end{equation*}
where $\text{D}_{\text{f}}\text{D}_{\text{b}}(\theta)$, $\text{D}_{\text{f}}\text{R}_{\text{b}}(\theta)$, $\text{D}_{\text{b}}\text{R}_{\text{f}}(\theta)$ and $\text{R}_{\text{f}}\text{R}_{\text{b}}(\theta)$ denote the normalized foreground-background, foreground-random, background-random and random-random pair counts at an angular separation of $\theta$. 

The cross-correlation measurements are also depicted in black in Figure \ref{xc_ac_data}, where they are compared with the ones used by \citet{GON21}, shown in gray\footnote{It should be noted that the measurements from \citet{GON21} were performed using the average-over-minitiles strategy.}. As expected, the cross-correlation signal is much weaker than the auto-correlation. The new data reach larger angular scales and have smaller statistical uncertainties due to the different and more efficient measurement methodology. As concluded in \citet{BON20}, these aspects are expected to lead to an improvement in the cosmological constraining power. It should be pointed out that unlike the data from \citet{GON21}, the cross-correlation signal at large angular scales does not seem to die off particularly steeply. This behavior (and its consequences on the cosmological estimates) are analyzed in Section \ref{sec:samplevariance}.

The covariance matrix is estimated for both observables through a Bootstrap method, which involves dividing the whole common area into $N$ subregions, resampling $N_r$ of them with replacement and repeating the process $N_b$ times. In essence, a random integer from 0 to $N$ is assigned to each subregion with the condition that all of them add up to $N_r$, effectively constructing a Bootstrap sample from the existing data on which one performs a measurement. This procedure is repeated $N_b$ times, each one with different assignments of random integers to each subregion. The covariance matrix is then computed via:

\begin{equation}
    \text{Cov}(\theta_i,\theta_j)=\frac{1}{N_b-1}\sum_{k=1}^{N_b}\,\bigg[\hat{w}_k(\theta_i)-\bar{\hat{w}}(\theta_i)\bigg]\bigg[\hat{w}_k(\theta_j)-\bar{\hat{w}}(\theta_j)\bigg]\label{covariance},
\end{equation}

where $\hat{w}_k$ denotes the measured correlation function from the $k^{\text{th}}$ Bootstrap sample and $\bar{\hat{w}}$ is the corresponding average value over all Bootstrap samples.

Regarding the choice of $N_r$, that is, the number of subregions to be drawn with replacement for each Bootstrap sample, we follow the conclusions of \citet{NORBERG09} and let $N_r=3N$, for which they obtained a very good agreement between the Bootstrap errors and those derived with an external estimate. To reach a compromise between the largest scales probed and the fact that we have 13 data points, we chose $N=20$, that is, we divided up each independent field into five patches. The procedure was repeated $N_b=10000$ times. Lastly, it should be noted that our internal approach does not take super-sample covariance \citep{LAC17} into account, since our dominant source of uncertainty is currently the sampling variance between the different fields used for the measurements (see Section \ref{sec:samplevariance}).

\begin{table*}[t]
\caption{Parameter prior distributions and summarized posterior results from the MCMC runs on the cross-correlation function with uniform and Gaussian priors on the $\beta$ parameter.} 
\centering 
\begin{tabular}{c c c c c c c c c} 
\hline 
\hline \\[-1.2ex]
\multicolumn{1}{c}{} & \multicolumn{4}{c}{Uniform $\beta$} & \multicolumn{4}{c}{Gaussian $\beta$} \\ 
\cmidrule(rl){2-5} \cmidrule(rl){6-9}
Parameter&Prior&Mean & Mode & 68\% CI & Prior & Mean & Mode & 68\% CI \\ 
\hline 
\\[-1ex]
$\alpha$ & $\mathcal{U}[0.00,1.50]$ & $0.83$ & $0.64$ & $[0.45,1.24]$ & $\mathcal{U}[0.00,1.50]$ &  $0.72$ & $0.64$ & $[0.32,1.04]$\\     
$\log M_{\text{min}}$ & $\mathcal{U}[10.00,16.00]$ & $11.67$ &  $11.73$ & $[11.48,11.94]$ & $\mathcal{U}[10.00,16.00]$ & $11.47$ & $11.54$ & $[11.34,11.69]$\\ [0.3ex]  
$\log M_1$ & $\mathcal{U}[10.00,16.00]$ & $13.41$ & $13.57$ & $[12.60,14.23]$ & $\mathcal{U}[10.00,16.00]$& $13.04$ & $12.94$ & $[12.02,13.71]$ \\   
$\Omega_m$ & $\mathcal{U}[0.10,1.00]$ & $0.21$ & $0.20$ & $[0.16,0.24]$ & $\mathcal{U}[0.10,1.00]$ & 0.23 & $0.21$ & $[0.17,0.26]$ \\
$\sigma_8$ & $\mathcal{U}[0.60,1.20]$ & $0.95$ & $-$ & $[0.87,1.20]$ & $\mathcal{U}[0.60,1.20]$ & $0.79$ & $0.79$ & $[0.69,0.89]$ \\
$h$ & $\mathcal{U}[0.50,1.00]$ & $0.70$ & $0.65$ & $[0.50,0.75]$ & $\mathcal{U}[0.50,1.00]$ & $0.72$ & $-$ & $[0.50,0.80]$\\
$\beta$ & $\mathcal{U}[1.50,3.50]$ & $2.46$ & $2.16$ & $[1.81,2.86]$ & $\mathcal{N}[2.90,0.04]$ & $2.90$ & $2.91$ & $[2.86,2.94]$ \\
\\[-1ex]
\hline 
\hline 
\end{tabular} 
\label{table1}
\end{table*}

\subsection{Parameter estimation}

For the purposes of this paper, the free parameters we consider here are: $M_{\text{min}}$, $M_1$, $\alpha$, $\Omega_m$, $\sigma_8$, $h$ and $\beta$. A flat $\Lambda$CDM cosmology is assumed throughout the paper, with $\Omega_b$ and $n_s$ fixed to the latest \textit{Planck} values \citep{PLANCK20}. For the estimation procedure, a Bayesian statistical approach is followed, for which the sampling of the posterior distributions was carried out via an MCMC algorithm using the open-source \emph{emcee} software package \citep{FOREMAN13}, a Python-based implementation of the Goodman \& Weare affine invariant MCMC ensemble sampler \citep{GOODMAN10}. 

Two main cases are distinguished in the analysis. The first one deals only with the angular cross-correlation function and assesses the parameter constraints that can be derived from its observation alone. The corresponding log-likelihood function can be described as a multivariate Gaussian, that is:

\begin{align*}
    \log{\mathcal{L}_{\text{cross}}\,(\theta_1,\ldots,\theta_m)}=-\frac{1}{2}&\bigg[m\log{(2\pi)}+\log{|C_{\text{cross}}|}\,+\\    &+\overrightarrow{\varepsilon}_{\text{cross}}^{\text{T}}C_{\text{cross}}^{-1}\,\overrightarrow{\varepsilon}_{\text{cross}}\bigg],
\end{align*}
where 
$\overrightarrow{\epsilon}_{\text{cross}}\equiv [\varepsilon_{\text{cross}}(\theta_1),\ldots,\varepsilon_{\text{cross}}(\theta_m)]$,

\begin{equation*}
    \varepsilon_{\text{cross}}(\theta_i)\equiv w_{\text{fb}}(\theta_i)-\hat{w}_{\text{cross}}(\theta_i)\quad\quad \forall i\in\{1,\ldots,m\}
\end{equation*}
and $C_{\text{cross}}$ is the covariance matrix associated to the cross-correlation measurements, computed according to Eq. \eqref{covariance}.

The second main case deals with a joint analysis of the foreground-background cross-correlation and the foreground-foreground auto-correlation. The corresponding log-likelihood will be expressed as:

\begin{align*}
    \log{\mathcal{L}_{\text{auto+cross}}\,(\theta_1,\ldots,\theta_m)}=-\frac{1}{2}&\bigg[m\log{(2\pi)}+\log{|C|}\,+\\    &+\overrightarrow{\varepsilon}^{\text{T}}C^{-1}\,\overrightarrow{\varepsilon}\bigg],
\end{align*}
where 
$\overrightarrow{\epsilon}\equiv [\varepsilon_{\text{cross}}(\theta_1),\ldots,\varepsilon_{\text{cross}}(\theta_m),\varepsilon_{\text{auto}}(\theta_1),\ldots,\varepsilon_{\text{auto}}(\theta_m)]$,

\begin{equation*}
    \varepsilon_{\text{auto}}(\theta_i)\equiv w_{\text{ff}}(\theta_i)-\hat{w}_{\text{auto}}(\theta_i)\quad\quad \forall i\in\{1,\ldots,m\}
\end{equation*}
and $C$ is the full covariance matrix associated to the cross- and auto-correlation measurements, again computed according to Eq. \eqref{covariance}. 

\begin{figure*}[t]

\centering
\minipage{0.25\textwidth}
  \includegraphics[width=\linewidth]{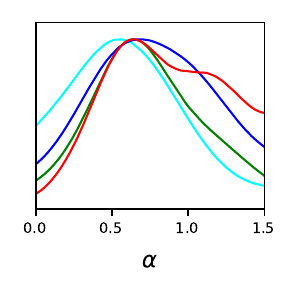}
\endminipage
\minipage{0.25\textwidth}
  \includegraphics[width=\linewidth]{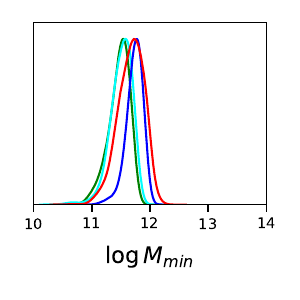}
\endminipage
\minipage{0.25\textwidth}
  \includegraphics[width=\linewidth]{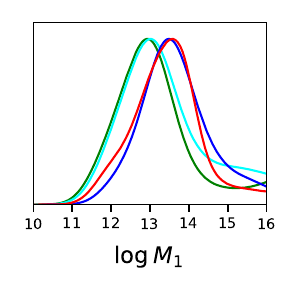}
\endminipage
\minipage{0.25\textwidth}
  \includegraphics[width=\linewidth]{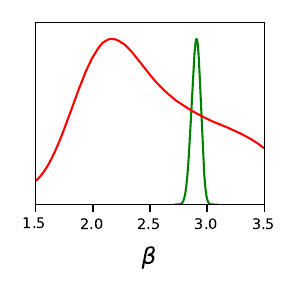}
\endminipage

\minipage{0.26\textwidth}
  \includegraphics[width=\linewidth]{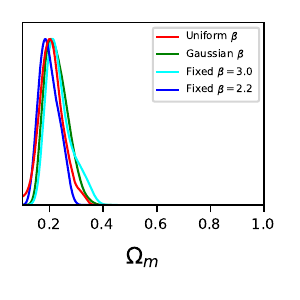}
\endminipage
\minipage{0.26\textwidth}
  \includegraphics[width=\linewidth]{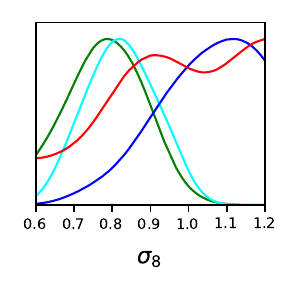}
\endminipage
\minipage{0.26\textwidth}
\includegraphics[width=\linewidth]{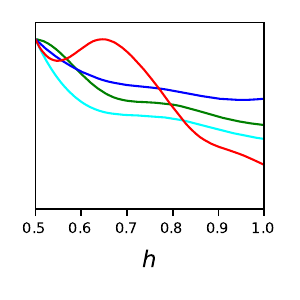}
\endminipage

\caption{Marginalized posterior distribution of the HOD, $\beta$ (top panels) and cosmological (bottom panels) parameters for the different cases discussed in Section 4.1.1, i.e., a uniform prior (in red), a Gaussian prior (in green), and fixed values on $\beta$ of 2.2. and 3.0 (in blue and cyan, respectively).}
\label{1D_xc_uniform_gaussian_fixed_beta}

\end{figure*}

\section{Results}

\subsection{Cross-correlation analysis}

\subsubsection{The $\beta$ parameter}
\label{sec:beta-section}

The preliminary cosmological and astrophysical results derived in \cite{BON20} and \cite{GON21} had assumed a fixed value of $\beta=3$, on account of the results found by \citet{LAPI11}, where the observed number counts of high-redshift \emph{Herschel} submillimeter galaxies had been successfully reproduced using an updated version of the galaxy formation model by \citet{LAPI06}. Indeed, high-redshift submillimeter galaxies were interpreted as massive protospheroids and the logarithmic slope of their intrinsic (unlensed) number counts at 350 $\mu$m was predicted to be near 3 at the 3$\sigma$ detection limit. However, the exact value of $\beta$ that should be used is not straightforward, since the behavior of the counts around the detection limit needs to be taken into account and the minimum flux that can be statistically boosted above the threshold cannot be predicted. Nonetheless, an analysis of the predicted integral number counts according to the above model allowed us to derive an average of the $\beta$ values in a sensible neighborhood of the detection limit. In essence, this translates into the possibility of considering a plausible prior for $\beta$ consisting of a Gaussian distribution with mean equal to 2.90 and standard deviation equal to 0.04.

Notwithstanding this analysis, we assess the importance of prior information on $\beta$ by studying four different cases in this first subsection assuming several choices for the $\beta$ parameter: a uniform prior distribution between 1.5 and 3.5, the aforementioned Gaussian prior distribution with mean 2.9 and standard deviation 0.04, and fixed values of 2.2 and 3.0. The last two cases are considered for different reasons: the first to assess large deviations from usual values and the second to make a comparison with previous results. 

However, as noted at the end of Section 2, the validity of the mean-redshift approximation needs to be tested before it can be adopted for all MCMC runs. We started by carrying out an analysis with the most general case (uniform prior distribution on $\beta$) to assess if there could be any important deviation from the full model. The resulting corner plot is depicted in Fig. \ref{corner_approx} and the summarized statistical results are shown in Table \ref{table_approx}. As can be clearly seen, only minor differences are present and we can safely adopt it for the rest of the paper. Moreover, this approximation is even more accurate in the tomographic setup of Paper II, where the model is evaluated within narrower redshift bins.
 
With regard to the four main MCMC runs of this first subsection, the corresponding full corner plots are shown in Figure \ref{corner_xc_uniform_gaussian_fixed_beta} in red, green, blue, and cyan for the cases of a uniform, Gaussian, and fixed 2.2. and 3.0 values for the $\beta$ priors, respectively, and the numerical results are summarized\footnote{Throughout the paper, all halo masses are expressed in $M_{\odot}/h$.} in Tables \ref{table1} and \ref{table2}. For visual purposes, the marginalized posterior distributions of all parameters are depicted in Figure \ref{1D_xc_uniform_gaussian_fixed_beta}.

A preliminary global view shows, as expected, that the case with the weakest priors (uniform distribution on $\beta$) yields the least stringent constraints on most parameters. However, the marginalized distributions of $M_{\text{min}}$, $M_1$, and $\sigma_8$ have a distinctive feature which sheds light onto the influence of the $\beta$ parameter; indeed, two different peak values for these parameters appear according to the two seemingly plausible values of $\beta$, which are around 2.2 and a 3.0. In particular, for $\sigma_8$, the interplay with the $\beta$ parameter becomes clear on the $\sigma_8-\beta$ plane, where larger values of the former imply smaller values of the latter. 

\begin{figure*}[t]
    \centering
    \includegraphics[width=0.72\textwidth]{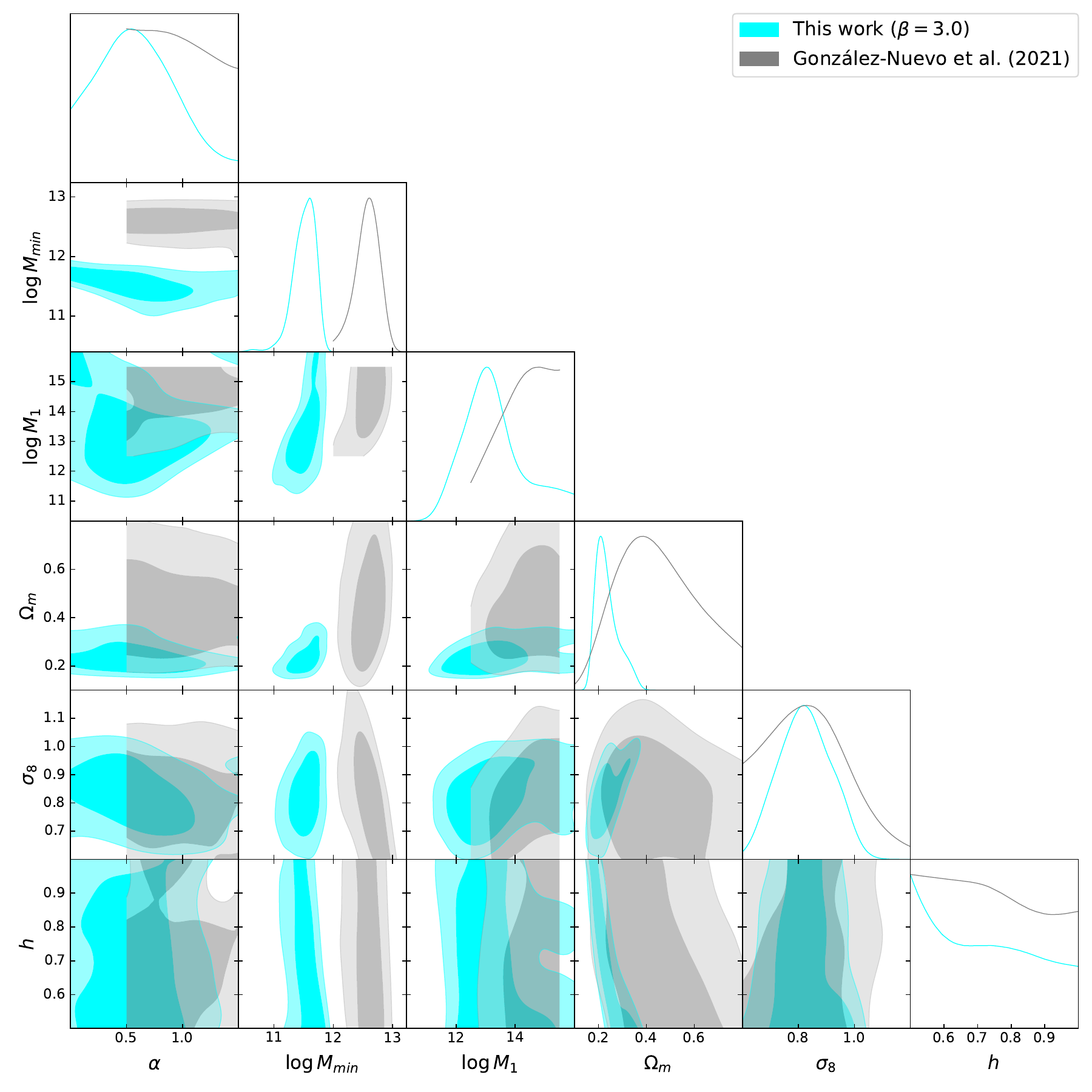}
    \caption{Marginalized posterior distributions and probability contours for the MCMC run on the cross-correlation function with a fixed value of $\beta=3.0$ (in cyan), compared to the results from \citet{GON21} depicted in gray. }
    \label{corner_xc_thiswork_vs_GN21}
\end{figure*}

Regarding the constraints on the HOD parameters, the least informative case (uniform prior on $\beta$) yields mean values of $\alpha=0.83$, $\log M_{\text{min}}=11.67$ and $\log M_1=13.41$, with 68\% credible intervals of $[0.45,1.24]$, $[11.48,11.94],$ and $[12.60,14.23]$. As we discuss below, even in this case, which has an additional degree of freedom with respect to \citet{GON21}, a remarkable improvement takes place concerning the determination of $\alpha$, which had not even been constrained, and $M_1$, for which mostly lower limits had been derived up to now. This is a direct consequence of the reduction of the error bars on small scales with respect to the measurement strategy in previous works. Nevertheless, the uncertainty in $M_{\text{min}}$ is comparable to that of \citet{GON21} due to the range of $\beta$ values that are allowed in the present work. Indeed (and interestingly), the posterior distribution of $\beta$ has a clear maximum at 2.20, but displays a long tail toward the high end. However, in the case of a narrow Gaussian prior around 2.90, the posterior does not deviate from it in any noticeable way. This fact, which in principle could be indicative of
a wrong assumption for the Gaussian $\beta$ prior, will be commented further in the next subsection.

\begin{figure*}[h]
\centering
\includegraphics[width=0.85\textwidth]{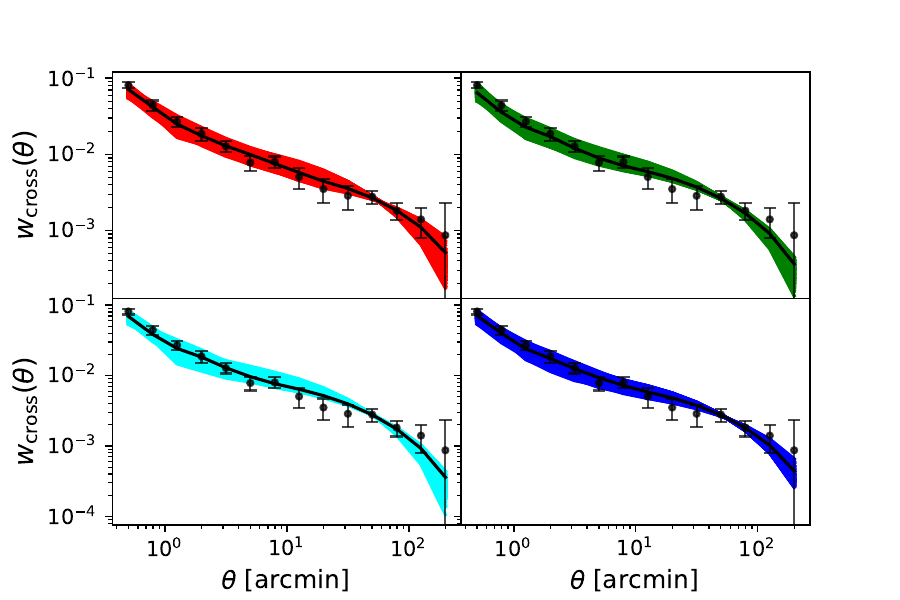}

\caption{Posterior-sampled angular cross-correlation function for the different cases of Subsection 4.1.1., that is, uniform prior on $\beta$ (in red), Gaussian prior on $\beta$ (in green) and fixed $\beta$ values of 3.0 and 2.2 (cyan and blue, respectively). The cross-correlation data are shown in black.}
\label{xc_sampling_uniform_gaussian_fixed_beta}
\end{figure*}

When the value of $\beta$ is fixed, either to 3.0. or 2.2, the constraints are clearly tightened, yielding mean values of $\log M_{\text{min}}=11.67^{+0.27}_{-0.19}$ and $11.47^{+0.22}_{-0.13}$ and $\log M_1=13.41^{+0.82}_{-0.81}$ and $13.04^{+0.67}_{-1.02}$, respectively. The difference between both cases explains the higher uncertainties discussed above. Indeed, as made clear in Figure \ref{corner_xc_uniform_gaussian_fixed_beta} by the red $\beta-\log M_{\text{min}}$ and $\beta-\log M_{1}$ contours and the marginalized posterior distributions of $M_{\text{min}}$ and $M_{1}$ for all cases, a lower (higher) value of $\beta$ is related to a higher (lower) value of both $M_{\text{min}}$ and $M_{1}$. This is a consequence of the interplay between halo masses and the logarithmic slope of the background number counts: the small-scale (1-halo) behavior of the cross-correlation caused by larger HOD masses can be counterbalanced by smaller $\beta$ values, which reduce the normalization of the signal.

Concerning cosmology, the case with the uniform prior on $\beta$ yields an unconstrained posterior distribution for the $\sigma_8$ parameter. However, when $\beta$ is fixed to 3.0 and 2.2, mean values of $\sigma_8=0.82^{+0.10}_{-0.10}$ and $1.02^{+0.18}_{-0.04}$ are obtained at 68\% credibility, respectively. With regard to the matter density parameter, it is the most robust in terms of dependence on the value of $\beta$. Compared to typical cosmological probes, low values of $\Omega_m$ are obtained in all four cases, with a mean of $0.21^{+0.03}_{-0.05}$ for the first (uniform prior on $\beta$),  $0.23^{+0.03}_{-0.06}$ for the second (Gaussian prior), $0.24^{+0.02}_{-0.06}$ for the third ($\beta=3.0$), and $0.20^{+0.03}_{-0.05}$ for the last ($\beta=2.2$). The Hubble constant, however, cannot be constrained at the moment in any case, as in previous studies \citep{BON20,GON21,BON21}. The reason behind this is twofold: firstly, the sensitivity of the angular cross-correlation function to the $h$ parameter is concentrated on the largest scales, where the uncertainties are still large. Secondly, given the degeneracy between $h$ and $\Omega_m$, which have opposite effects, and the fact that the angular cross-correlation is much more sensitive to the latter parameter, constraining the former seems to prove challenging at the current stage. 

Nonetheless, for comparison purposes, Figure \ref{corner_xc_thiswork_vs_GN21} depicts the results for the $\beta=3.0$ case from this paper to those of \citet{GON21}, which had used the same $\beta$ value. The results show a remarkable overall improvement in terms of parameter uncertainties, mainly due to the reduction of error bars and to the possibility of reaching larger angular scales with the new methodology. The differences in the specific posterior distributions can be explained by two different facts\footnote{It should also be noted that tighter prior distributions were imposed for the HOD parameters in \cite{GON21}.}. Firstly, and as discussed in Section 3.2., the data from \citet{GON21}, which employed a different measurement methodology, might have been biased due to the relative arbitrariness of the so-called integral constraint correction on large scales. Secondly, and as will be highlighted in the next subsection, the steeper ones fall above 20 arcmin in the data from \citet{GON21} can only be accounted for with larger values of $\Omega_m$.

Figure \ref{xc_sampling_uniform_gaussian_fixed_beta} shows the posterior sampling of the cross-correlation function for all four cases in red, green, blue and cyan, respectively, along with the best fit (solid black line) and the data. The model explains the data correctly on most scales, but it tends to slightly overestimate the signal between 10 and 40 arcmin and to underestimate it above 60 arcmin, especially when $\beta$ has larger values (2.9 and 3.0).  Moreover, the tight small-scale uncertainties of the data seem to induce a preference toward a steeper fall above 60 arcmin than the data suggest. Indeed (as commented in Section 3.2), our current data do not die off particularly steeply and a sensitivity analysis shows that this large-scale behavior forces particularly low values of $\Omega_m$. This issue will be studied in detail in the next subsection.

The first conclusion to be drawn up to now is that a priori knowledge on the $\beta$ parameter is a necessary condition for constraining cosmology using the submillimeter galaxy magnification bias with a single redshift bin. Indeed, although the matter density parameter is barely affected by its value, too loose a prior on $\beta$ erases the possibility to constrain $\sigma_8$ as a consequence of the large degeneracy and interplay between both parameters. Furthermore, a wrong assumption of a fixed value of $\beta$ can substantially bias the constraint on $\sigma_8$, with larger values of the former linked to lower values of the latter. In particular, it should be stressed once again that previous works on the submillimeter galaxy magnification bias assumed a value of $\beta=3.0$.

\subsubsection{Sampling variance and the large-scale cross-correlation}

\label{sec:samplevariance}

\begin{figure}[h]
    \centering
    \includegraphics[width=\columnwidth]{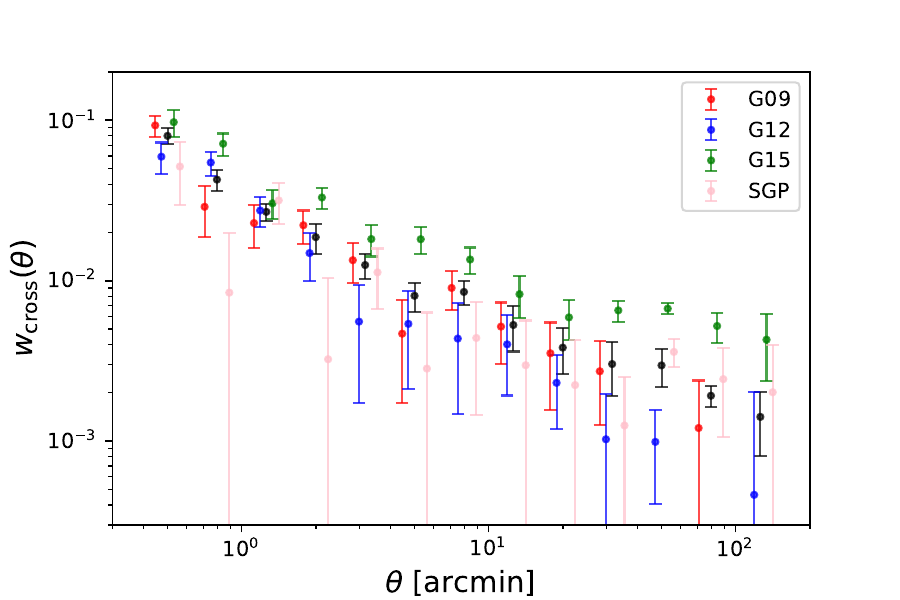}
    \caption{Measurements of the foreground-background angular  cross-correlation function in each independent field separately (G09, in red, G12, in blue, G15, in green and SGP, in pink), compared to the global measurement combining all fields (in black).}
    \label{xc_data_fields}
\end{figure}

The moderate fall in the large-scale behavior of the cross-correlation function seems to suggest a more thorough analysis. Given the fact that we have four spatially separated regions in the sky, we decided to measure the angular cross-correlation function in each of the different fields independently so as to examine if there are substantial differences among them. Figure \ref{xc_data_fields} shows the corresponding results for the G09, G12, G15 and SGP regions (in red, blue, green and pink, respectively) along with the overall data (in black). 

Although the cross-correlation signal shows a relatively stable profile in the case where all regions are combined, the data display non-negligible variations among the fields that are nonetheless mostly contained within the error bars. However the G15 regions represents a clear exception, with a systematically higher signal on all scales. Regarding possible explanations of this phenomenon, it should first be noted that the selection criteria of both the foreground and background samples are uniform across all four regions and the redshift distributions barely differ. Moreover, stellar masses derived via stellar population fitting of the SEDs as found in the GAMA catalogs show a remarkable uniformity among the fields.

Another possibility would be failing to correctly consider the selection function of the samples in the construction of the random catalogs. This is automatically taken into account for the foreground sample, since we made use of the random catalogs generated by the GAMA team and available on the database that reproduce the underlying selection function\footnote{Using a purely Poissonian random catalog was shown to introduce almost negligible variations in the signal.}. As for the background sample, the spatial variation in the instrumental noise is considered via H-ATLAS noise maps, which are imprinted in the random catalog. Even if this didn't account for the whole selection function of the background sample, the differences among regions are already present in the foreground sample itself, so they should not be sourced by a missing consideration in the selection function.

\begin{figure}[h]
    \centering
    \includegraphics[width=\columnwidth]{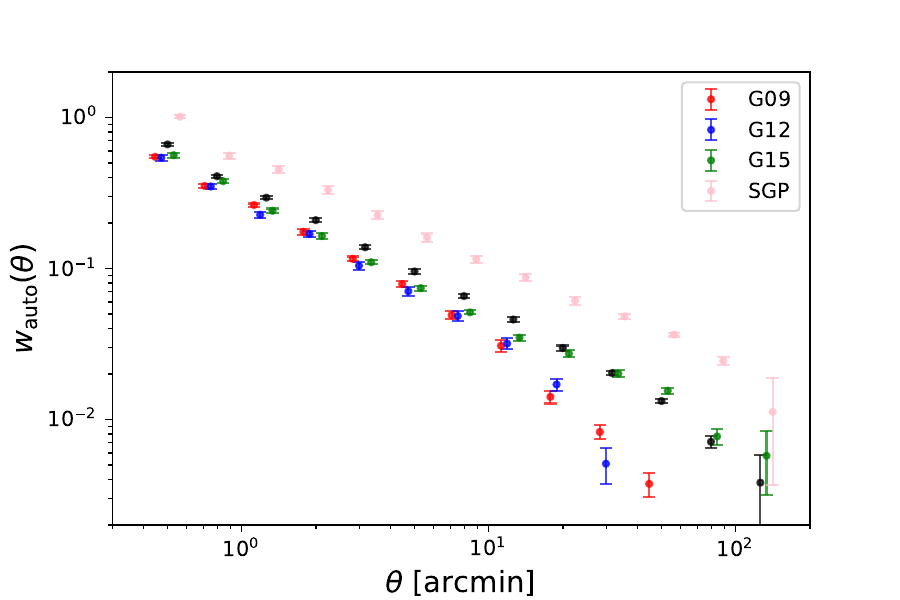}
    \caption{Measurements of the foreground angular auto-correlation function in each independent field separately (G09, in red, G12, in blue, G15, in green and SGP, in pink), compared to the global measurement combining all fields (in black).}
    \label{ac_data_fields}
\end{figure}

Indeed, this variability is observed even more strongly in the angular auto-correlation function of the foreground sample, depicted in Figure \ref{ac_data_fields} with the same colors as the cross-correlation. The foreground galaxies in the G15 region do not seem to cluster particularly strongly (which would explain a higher cross-correlation signal), but the large variation among fields, even at small scales, seems to point to sampling variance as the underlying cause. Although this issue will be studied further in future work with a larger data set, we performed a preliminary analysis by excluding the seemingly anomalous G15 region.

Figure \ref{xc_ac_data} shows a comparison between the cross-correlation measurements taking all four regions into account (in black) and the corresponding ones resulting from excluding the G15 region (in olive green). Indeed, the removal of G15 induces a steeper fall above 40 arcmin, more in keeping with the qualitative behavior of the data from \citet{GON21}. Although this region was also included in the cross-correlation computation in \citet{GON21}, we believe that the different measurement strategy (involving splitting each field into 16 subregions and averaging the cross-correlation signal) might have helped mitigate the differences.

In order to assess the influence of the (unexpectedly) moderate fall in the large-scale data, we considered
excluding the G15 region from the computation of the angular cross-correlation function.  We performed two MCMC runs, namely, with a uniform and a Gaussian prior on $\beta$, and compared the results with those from the previous subsection. The full cornerplots are shown in Figs. \ref{corner_xc_uniform_all_vs_noG15} and \ref{corner_xc_gaussian_all_vs_noG15}, respectively. 

\begin{figure}[h]

\centering

\minipage{0.5\columnwidth}
  \includegraphics[width=\linewidth]{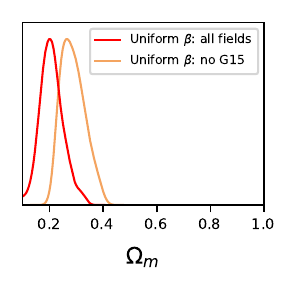}
\endminipage
\minipage{0.5\columnwidth}
  \includegraphics[width=\linewidth]{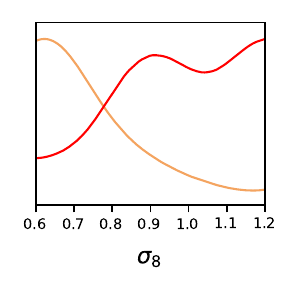}
\endminipage

\minipage{0.5\columnwidth}
  \includegraphics[width=\linewidth]{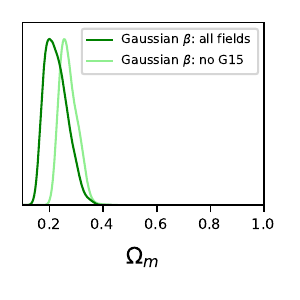}
\endminipage
\minipage{0.5\columnwidth}
  \includegraphics[width=\linewidth]{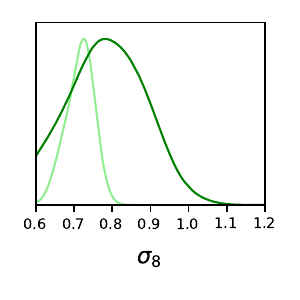}
\endminipage

\caption{Marginalized posterior distribution for the $\Omega_m$ and $\sigma_8$ parameters. Top panel: Results for the MCMC runs with a uniform prior on $\beta$. Bottom panel: Results for the MCMC runs with a Gaussian prior on $\beta$.}
\label{1D_omgM_and_sgm8_noG15_and_nolargescale}
\end{figure}

The results on the HOD parameters behave as expected for both cases. Indeed, the exclusion of the G15 region reduces the cross-correlation signal even at small scales, which implies smaller galaxy halo masses. As regards cosmology, important differences appear with respect to using all fields. Fig. \ref{1D_omgM_and_sgm8_noG15_and_nolargescale} summarizes the deviations for $\Omega_m$ and $\sigma_8$. For both cases, $\Omega_m$ is the most affected parameter, with the same qualitative behavior: the exclusion of the G15 region implies higher values of $\Omega_m$ due to the steeper fall at the large-scale end. In particular, for a uniform (Gaussian) prior distribution on $\beta$, the mean value goes from $\Omega_m=0.21^{+0.03}_{-0.05}$ ($0.23^{+0.03}_{-0.06}$) using all data and regions to $\Omega_m=0.29^{+0.03}_{-0.06}$ ($0.27^{+0.03}_{-0.04}$) when the G15 region is excluded. As regards $\sigma_8$, the interplay with the $\beta$ parameter still doesn't allow a clear determination when no information about the latter is given, but an upper limit of $\sigma_8<0.81$ is obtained at 68\% credibility. However, for the Gaussian prior on $\beta$, the  distribution is tightened, with the mean value going from $\sigma_8=0.79^{+0.10}_{-0.10}$ using all regions to $\sigma_8=0.72^{+0.04}_{-0.04}$ when the G15 region is excluded. Interestingly, the posterior distribution of the Hubble constant, although wide, displays (for the first time in submillimeter galaxy magnification bias studies) a clear maximum and we derived a mean value of $h=0.79^{+0.13}_{-0.14}$. It should be noticed that, although the observed anticorrelation between $\Omega_m$ and $h$ would be expected to imply smaller values of the Hubble constant with respect to the previous case (given the larger $\Omega_m$ after removing the G15 region),  such values are not favored. This seems to be a consequence of the weaker overall signal compared to the four-field case. Indeed, even if the anticorrelation between the two parameters is still present, too small a value for the Hubble constant would not allow $\Omega_m$ (which cannot be arbitrarily high due to the interplay with $\sigma_8$ and the HOD parameters) to explain the weaker signal\footnote{A sensitivity analysis shows that, unlike for the Hubble constant, variations of $\Omega_m$ still affect the signal at small and intermediate scales and thus imply a non-trivial interplay with the HOD parameters.}. Lastly, and as opposed to the previous case, the marginalized distribution of $\beta$ for the uniform-prior case now displays a clear maximum at 3, well above the value of 2.2. Therefore, the physically motivated $\beta\sim 2.9$ is recovered by excluding the G15 region, effectively solving the apparent inconsistency described in the previous subsection.

The conclusion at this point is that sampling variance can induce a bias in the cosmological parameter constraints derived from an analysis of the submillimeter galaxy magnification bias. In our case, an excess of cross-correlation seems to be present in the G15 region. When excluded, the behavior resembles the qualitative fall of the signal from \citet{GON21}, which averaged over smaller subregions and was thus less likely to be affected by large-scale inhomogeneities. Further work is needed along this line to confirm this hypothesis by enlarging the galaxy samples.

\subsection{Cross- and auto-correlation joint analysis}

Inspired by studies of galaxy-galaxy lensing, where the clustering of galaxies is added to tighten constraints \citep{ABBOTT18,ABBOTT22}, we performed a joint analysis for both the angular cross- and foreground auto-correlation functions. To assess whether galaxy clustering could help tighten the constraints, we analyzed our base case: a Gaussian prior for $\beta$. The full cornerplot with the results is shown in Fig. \ref{corner_xc_ac_gaussian} in purple, where the cross-correlation-only case is also depicted in green for comparison purposes. Table \ref{table5} presents the corresponding summarized statistical results.

The constraints on all parameters are extremely tightened after the incorporation of the auto-correlation function. Regarding the HOD, we obtained mean values of $\alpha=0.92^{+0.04}_{-0.05}$, $\log M_{\text{min}}=11.54^{+0.09}_{-0.11}$ and $\log M_1=12.41^{+0.25}_{-0.17}$. Interestingly, the distribution of the matter density parameter is displaced
toward higher values, with a mean value of $\Omega_m=0.36^{+0.01}_{-0.02}$. The normalization parameter of the power spectrum is reduced to lower values with respect to the sole cross-correlation measurement, with a mean of $\sigma_8=0.72^{+0.04}_{-0.03}$. Regarding the Hubble constant, we obtained an extremely saturated distribution toward low values, possibly related to the apparent inconsistency between both observables.

\begin{figure}[h]
    \centering
    \includegraphics[width=\columnwidth]{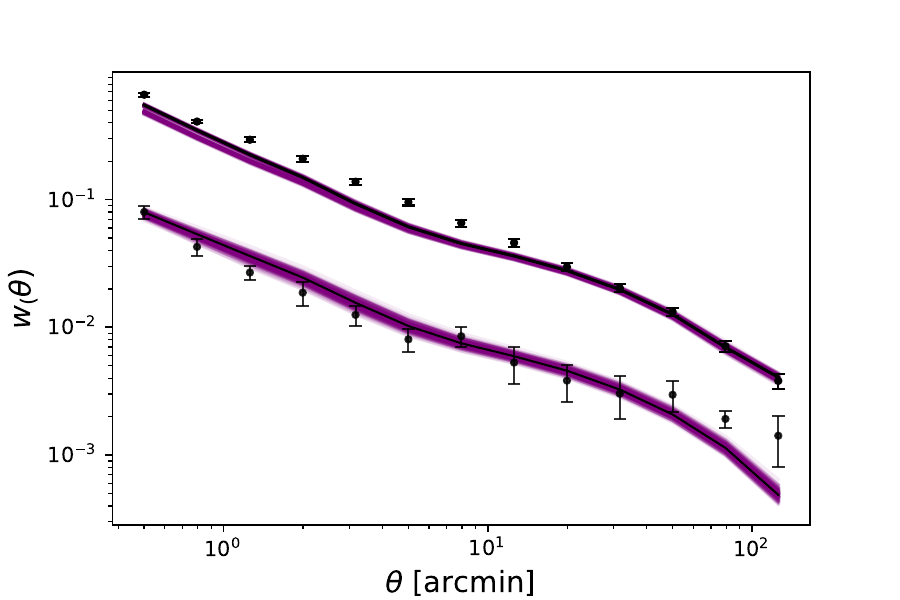}
    \caption{Posterior-sampled angular auto- and cross-correlation functions from the joint analysis of both observables. The data are shown in black.}
    \label{xc_ac_sampling}
\end{figure}

Indeed, the sampling of the posterior distribution, shown in Fig. \ref{xc_ac_sampling}, reveals a poor joint fit to the data, mainly due to the underestimation of the auto-correlation signal below 20 arcmin. A similar issue was found in the joint clustering and galaxy-galaxy lensing analysis of \citet{ABBOTT22}, where an internal inconsistency between the tangential shear and the auto-correlation function of the redMaGiC lens sample was apparent. Although sampling variance would again be a plausible explanation in our case and, possibly, the most relevant one at the current stage, other reasons could be behind this discrepancy. Indeed, as shown by hydrodynamical simulations, baryonic feedback can impact the galaxy and halo distributions differently \citep{vanDaalen2014,RENNEBY20,vanDaalen2020}, which can effectively reduce the lensing signal in the one-halo regime while having no impact on clustering \citep{AMON23}. Moreover, assembly bias (i.e., the dependency of halo occupation and clustering on secondary properties other than halo mass) can suppress the lensing signal on small and even intermediate scales ($\lesssim 5-10$ Mpc/$h$), as shown by \citet{GAO05}, \citet{WECHSLER06}, \citet{LEAUTHAUD17}, and \citet{AMON23}, among others. This matter goes beyond the scope of this paper and will be pursued in future work.

However, slightly different HOD values were found in previous related works \citep{BON20,BON21} from a separated analysis of the auto- and cross-correlation functions. Since this could be related to the above discussion, we decided to run a preliminary joint test allowing a different HOD model for each observable to analyze if the galaxy-halo connection could respond differently to each physical situation. The results are shown in Fig. \ref{xc_ac_cornerplot_dHOD} and Table \ref{table6}. The overall fit, shown in Fig. \ref{xc_ac_sampling_dHOD}, is better than the previous case, but the cross-correlation is now overestimated below 30 arcmin. The galaxy-halo connection, although compatible at $2\sigma$, differs slightly between the auto- and cross-correlation models, since $\log M_{\text{min}}^{\text{auto}}=11.58^{+0.12}_{-0.22}$, $\log M_{\text{min}}^{\text{cross}}=11.83^{+0.08}_{-0.08}$ , $\log M_1^{\text{auto}}=12.33^{+0.24}_{-0.47}$ (at 68\% credibility), and $\log M_1^{\text{cross}}>12.34$ (at 95\% credibility). In turn, this implies a very high value of $\sigma_8$, whose mean value is constrained to be $\sigma_8=1.10^{+0.05}_{-0.04}$. We found a matter density parameter that is perfectly consistent with the previous cases ($\Omega_m=0.34^{+0.05}_{-0.05}$) and, interestingly, the Hubble constant is no longer unconstrained, with a mean value of $h=0.69^{+0.06}_{-0.17}$. These results should of course be taken with a pinch of salt and further work will be needed regarding the enlargement of the galaxy sample to shed light on this apparent inconsistency.

\section{Summary and conclusions}

This paper has addressed the cosmological constraints resulting from the refinement of the methodology of the submillimeter galaxy magnification bias. With an improved strategy with respect to previous works, we have measured the weak-lensing induced angular cross-correlation function between a sample of H-ATLAS submillimeter galaxies with photometric redshifts $1.2<z<4.0$ and a sample of GAMA II 
galaxies with spectroscopic redshifts $0.2<z<0.8$ using a single wide foreground bin. Additional aspects on the modeling side have been addressed, as the relevance of the logarithmic slope of the background number counts and the explicit numerical computation of the two-halo terms of the halo model power spectra. Additionally, we have measured the angular auto-correlation function of the GAMA II galaxies to assess the possibility of improving the constraints on cosmology through additional information about the foreground sample. By means of a halo model interpretation of the galaxy-matter and galaxy-galaxy power spectra, we carried out a Bayesian analysis through an MCMC algorithm to obtain posterior probability distributions about the HOD and the cosmological parameters $\Omega_m$, $\sigma_8$ and $h$ of a flat $\Lambda$CDM model. 

We began the analysis using only the cross-correlation data and an important point was immediately raised: the value of $\beta$ (i.e., the logarithmic slope of the background (unlensed) integral number counts) can have a large influence on the constraints of some of the parameters, namely $M_{\text{min}}$, $M_1$ and, most importantly, $\sigma_8$. Although the matter density parameter, $\Omega_m$, was barely affected by this issue, we conclude that a priori information on the value of $\beta$ is of paramount importance to derive unbiased constraints using the cross-correlation data alone. Indeed, prior studies relied on a fixed value of $\beta=3$, but the degeneracy and interplay between this parameter and $\sigma_8$ imply that larger values of the former are related to lower values of the latter. Since the choice of a uniform prior distribution for $\beta$ does not produce any constraint on $\sigma_8$, an analysis of the (predicted) intrinsic integral number counts allowed us to restrict the prior distribution of $\beta$ on physically-motivated grounds: $\beta=2.90\pm0.04$. For this case, we obtained mean values of $\Omega_m=0.23^{+0.03}_{-0.06}$ and $\sigma_8=0.79^{+0.10}_{-0.10}$ at 68\% credibility and no deviation whatsoever from its prior distribution for $\beta$.

However, inspecting the sampling of the posterior distribution and comparing the current cross-correlation data with the ones from \citet{GON21} raised suspicion that there could be an intrinsic bias in the signal. Indeed, the moderate large-scale fall seems to be induced by sampling variance; more precisely, by the seemingly anomalous behavior of the G15 region, where the signal is notably stronger than the rest at all scales. To test how this could influence the results, we performed additional MCMC runs by excluding the G15 region from the measurement of the signal. The two cases that we studied (uniform and Gaussian prior on $\beta$) yielded qualitatively similar results, confirming that nonnegligible differences arose with respect to the previous case. Although the influence of $\beta$ on the constraining power over the $\sigma_8$ parameter remained, the exclusion of the G15 region yielded a mean value of $\Omega_m=0.29^{+0.03}_{-0.06}$ for the matter density parameter. When a Gaussian prior was assigned to $\beta$, we obtained mean values of $\Omega_m=0.27^{+0.02}_{-0.04}$ and $\sigma_8=0.72^{+0.04}_{-0.04}$, larger and smaller than the results of the previous paragraph, where all regions had been used. Interestingly, in this case we obtained the first (albeit loose) constraint on then the Hubble constant using the submillimeter galaxy magnification bias, for which  results up to now had only consisted in one-sided limits. A mean value of $h=0.79^{+0.13}_{-0.14}$ was obtained at 68\% credibility.

When a joint analysis of the angular cross- and foreground auto-correlation functions is carried out, parameter constraints are tightened with respect to the previous case, but the fit is worsened due to a clear underestimation of the auto-correlation signal that is reminiscent of the internal inconsistency found in \citet{ABBOTT22} between tangential shear and autocorrelation measurements in one of their lens samples. Although sampling variance is still likely the most plausible explanation, other phenomena, such as baryonic feedback or assembly bias acting on small to intermediate scales, may be behind the apparent discrepancy between both observables. 

Along with the inclusion of these smaller-scale effects, the overall behavior of the G15 region (and, more generally, the issue of sampling variance) will be a subject for future studies, since there is no physical indication that it should be discarded for our analyses. The addition of more independent regions coming from other surveys will allow us to estimate the significance of this deviation of the cross-correlation signal by performing a statistical analysis. 

\begin{acknowledgements}
LB, JGN, JMC and DC acknowledge the PID2021-125630NB-I00 project funded by MCIN/AEI/10.13039/501100011033/FEDER, UE. LB also acknowledges the CNS2022-135748 project funded by MCIN/AEI/10.13039/501100011033 and by the EU “NextGenerationEU/PRTR”. JMC also acknowledges financial support from the SV-PA-21-AYUD/2021/51301 project.\\
We deeply acknowledge the CINECA award under the ISCRA initiative, for the availability of high performance computing resources and support. In particular the projects `SIS22\_lapi', `SIS23\_lapi' in the framework `Convenzione triennale SISSA-CINECA'.\\
The Herschel-ATLAS is a project with Herschel, which is an ESA space observatory with science instruments provided by European-led Principal Investigator consortia and with important participation from NASA. The H-ATLAS web- site is http://www.h-atlas.org. GAMA is a joint European- Australasian project based around a spectroscopic campaign using the Anglo- Australian Telescope. The GAMA input catalogue is based on data taken from the Sloan Digital Sky Survey and the UKIRT Infrared Deep Sky Survey. Complementary imaging of the GAMA regions is being obtained by a number of independent survey programs including GALEX MIS, VST KIDS, VISTA VIKING, WISE, Herschel-ATLAS, GMRT and ASKAP providing UV to radio coverage. GAMA is funded by the STFC (UK), the ARC (Australia), the AAO, and the participating institutions. The GAMA web- site is: http://www.gama-survey.org/.\\
This research has made use of the python packages \texttt{ipython} \citep{ipython}, \texttt{matplotlib} \citep{matplotlib} and \texttt{Scipy} \citep{scipy}.
\end{acknowledgements}

\bibliographystyle{aa} 
\bibliography{main} 

\appendix

\section{Ingredients of the halo model}

This section describes in detail the computation of the halo model prescription for the galaxy and galaxy-matter power spectra. The halo mass function is expressed as:
\begin{equation*}
    n(M,z)=\frac{\rho_0}{M^2}\,f_{\text{ST}}(\nu)\bigg|\frac{d\log \nu}{d\log M}\bigg|,
\end{equation*}
where 
\begin{equation*}
    f_{\text{ST}}(\nu)=A\sqrt{\frac{a\nu}{2\pi}}\bigg[1+\bigg(\frac{1}{a\nu}\bigg)^p\bigg]e^{-a\nu/2}
\end{equation*}
is the Sheth and Tormen model \citep{ST99}, for which $A=0.33$, $a=0.75,$ and $p=0.3$. The $\nu$ parameter is defined as
\begin{equation*}
    \nu(M,z)\equiv \bigg[\frac{\hat{\delta}_c(z)}{\sigma(M,z)}\bigg]^2,
\end{equation*}
where $\hat{\delta}_c(z)$ is linear critical overdensity at redshift $z$ for a region to collapse into a halo at this redshift \citep[computed via the fit of][] {KITAYAMA96} and $\sigma(M,z)\equiv D(z)\sigma(M,0)$, where $D(z)$ is the linear growth factor for a $\Lambda$CDM universe (normalized at $z=0$) and $\sigma(M,0)$ is the square root of the mass variance of the filtered linear overdensity field at $z=0$.

The linear deterministic halo bias is computed from the above halo mass function via the peak-background split \citep{ST99}, yielding
\begin{equation*}
    b_1(M,z)=1+\frac{a\nu-1}{\hat{\delta}_c(z)}+\frac{2p/\hat{\delta}_c(z)}{1+(a\nu)^p}.
\end{equation*}
Moreover, the mean number of galaxies in a halo of mass $M$, $\langle N\rangle_M$, follows the model described in Eq. \eqref{HOD} and the mean number density of galaxies at redshift, $z,$ is computed via:

\begin{equation*}
    \bar{n}_g(z)=\int_0^{\infty}dM\,n(M,z)\,\langle N\rangle_M.
\end{equation*}

The matter transfer function used to compute the linear matter power spectrum is computed via the analytical formula of \cite{EISENSTEIN98}, which takes baryonic effects into account. This approach has been favored over a numerical one based on Boltzmann codes due to computation time, since no significant differences were found for our purposes.

Lastly, the halo density profile is taken to follow the two-parameter NFW model \citep{NAVARRO97}, that is,
\begin{equation*}
    \rho(r)=\frac{\rho_s}{(r/r_s)(1+r/r_s)^2}.
\end{equation*}
The halo is effectively truncated at a comoving virial radius, $r_h$, which yields a relation between $\rho_s$ and the mass $M$ of the virialized halo (or, equivalently, its mean density, $\rho_h$):
\begin{equation*}
    \rho_s=\frac{\rho_h}{3}\frac{c}{\ln{(1+c)}-c/1+c},
\end{equation*}
where $c\equiv r_h/r_s$ is the halo concentration parameter. Since the mass of the halo satisfies
\begin{equation*}
    M=\frac{4}{3}\,\pi \,r^3_h(z)\, \rho_h(z),
\end{equation*}
the choice of both a typical density for a collapsed halo, $\rho_h$, and of a prescription for the mean mass-concentration relation, $c(M,z)$, completely specifies the values of $r_s$ and $\rho_s$; thus, the NFW profile of a typical halo of mass, $M,$ at redshift, $z$. We have chosen a virial overdensity criterion, that is, 
\begin{equation*}
    \rho_h(z)=\Delta_{\text{vir}}(z)\,\bar{\rho}_0,
\end{equation*}
where $\Delta_{\text{vir}}(z)$ has been computed through the fit by \cite{WEINBERG03}. Regarding the mass-concentration relation, we have used that of \cite{BULLOCK01}. With all this in mind, the normalized Fourier transform of a typical halo of mass, $M,$ at redshift, $z,$ is expressed as:\ 
\begin{align*}
    u(k|M,z)&=\frac{1}{\ln{(1+c)}-c/1+c}\Big[\sin{kr_s}\big[\text{Si}([1+c]kr_s)-\text{Si}(kr_s)\big]\\
    &-\frac{\sin{ckr_s}}{[1+c]kr_s}+\cos{kr_s}\big[\text{Ci}([1+c]kr_s)-\text{Ci}(kr_s)\big]\Big],
\end{align*}
where Si and Ci are the sine and cosine integrals, respectively.

\section{Numerical issue about the computation of the two-halo term}

As pointed out by \cite{MEAD20} and \cite{MEAD21}, the evaluation of the two-halo terms of the power spectra involving the matter field presents a numerical problem. The halo model considers that all matter is bound up into distinct halos, which is translated into the following condition for the halo mass function:
\begin{equation*}
    \int_0^{\infty}dM\,\frac{n(M,z)}{\bar{\rho}_0}\,M=1,
\end{equation*}
which, in the case of the Sheth \& Tormen model, fixes the normalization parameter $A$ in terms of $p$. However, an additional condition involving the linear halo bias must be enforced in order for the halo model to reproduce the correct large-scale behavior of the power spectra. This condition is expressed as:
\begin{equation*}
    \int_{0}^{\infty} dM\,\frac{n(M,z)}{\bar{\rho}_0}\,b_1(M,z)\,M=1
\end{equation*}
and presents a very relevant numerical problem. Indeed, defining
\begin{equation}
    I(M_a,M_b,k,z)\equiv \int_{M_a}^{M_b}\,dM\,\frac{n(M,z)}{\bar{\rho}_0}\,b_1(M,z)\,M\,u(k|M,z)\label{integral},
\end{equation}
the halo model requires that $I(M_a\to0,M_b\to\infty,k\to0,z)\to 1$. However, choosing sensible limits of $M_a=10^{8}M_{\odot}/h$ and $M_b=10^{8}M_{\odot}/h$, a regular \textit{Planck} cosmology yields a value of $\approx 0.7$. The problem stems from the fact that typical models for the halo mass function assign a large fraction of the total matter to low-mass halos, causing convergence of the above integral to be extremely slow as the lower limit of the integral becomes smaller.

\begin{figure}[h]
    \centering
    \includegraphics[width=0.5\textwidth]{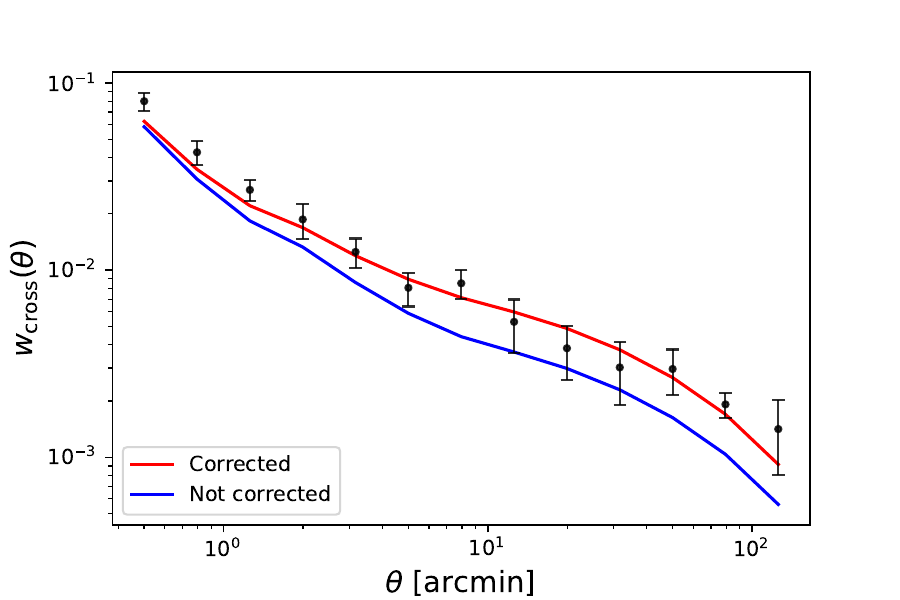}
    \caption{Effect on the cross-correlation function of the numerical correction in the computation of the two-halo term of the galaxy-matter power spectrum}.
    \label{2halocorrection}
\end{figure}

To work around the problem, \cite{SCHMIDT16} suggests modifying the halo mass function under the assumption that all mass below $M_a$ is contained in halos of a mass that is exactly $M_a$. This amounts to making the substitution: 
\begin{equation*}
    n(M,z)\Longrightarrow n(M,z)\Theta(M-M_a)+\frac{A(M_a)}{b_1(M_a)M_a/\bar{\rho}_0}\delta_{\text{D}}(M-M_a)
\end{equation*}
in \eqref{integral}, where
\begin{equation*}
    A(M_a)\equiv 1-\int_{M_a}^{\infty}dM \frac{n(M,z)}{\bar{\rho}_0}\,b_1(M,z)\,M.
\end{equation*}

If this is not taken into account, an extremely strong bias is induced in any cross-correlation involving the matter field. For instance, Fig. \ref{2halocorrection} shows the effect of not including this correction in the angular cross-correlation function. The discrepancy increases with the angular scale, reaching about 60\% above 5 arcmin and compromising any reliable constraint on cosmology. We stress that this should be addressed in any computation of halo model cross-correlations involving matter.

\section{ Integral constraint correction}

Given a certain patch, the number of detected galaxies in it will certainly be higher or lower than
what we would have in a fair sample of the Universe, thus affecting our estimates in the random
catalog of the patch. Averaging over a large number of patches, as done in previous works \citep{BON20,BON21,CUE21,CUE22}, tends to introduce an artificial weakening of the observed
clustering signal.  This is because sources sufficiently close to edges of the corresponding field are less likely to have
pairs at large distances. This effectively causes the estimated cross-correlation to be biased low by a
constant \citep{ADELBERGER05}, so that:
\begin{equation*}
    \hat{w}_{\text{ideal}}(\theta)=\hat{w}(\theta)+\text{IC.}
\end{equation*}

Although there are theoretical approaches to estimate the IC for a particular scanning strategy \citep[see e.g.,][]{ADELBERGER05}, in practice, it is commonly estimated numerically using random-random counts. Specifically, the IC can be estimated for the cross-correlation using the formula:

\begin{equation}
\text{IC}=\frac{\sum_i{\rm{R}_{\text{f}}\rm{R}_{\text{b}}(\theta_i)w_{\text{ideal}}(\theta_i)}}{\sum_i{\rm{R}_{\text{f}}\rm{R}_{\text{b}}(\theta_i)}},
\end{equation}
where $w_{\text{ideal}}(\theta)$ is an assumed model for the cross-correlation function. An equivalent expression is used for the auto-correlation.

However, the obvious caveat is that the ideal model for the cross-correlation function is not known. Two approaches appear to be reasonable in this situation. Thus, inspired by the usual procedure of the angular auto-correlation estimation, we can first assume a power-law model of $w_{\text{ideal}}(\theta)= A \theta^{-\gamma}$. The procedure, however, is highly dependent on the limiting angular scales used for the fit. Although we can derive a wide-range estimate of IC$\sim 7-18\times10^{-4}$, we prefer to abandon this method in favor of a more precise alternative that does still make use of this value.

The second approach implies assuming the halo model as the ideal cross-correlation. An additional problem stems from the fact that the correction would depend on the HOD and, most importantly, on cosmology. This is crucial, because the largest angular scales are the most important for constraining cosmological parameters but, at the same time, they are the most affected by the adopted IC correction. Therefore, we performed the following analysis to try and refine the procedure: firstly, we performed maximum likelihood estimation searches for the HOD parameters using only the measurements in the one-halo regime and random uniform cosmological parameters. With the obtained HOD parameters, we computed the IC distribution for 100 random sets of cosmological parameters. We then discarded those cosmologies resulting in IC values outside the range $7-18\times 10^{-4}$, as obtained above. The final estimate we obtained was IC=$(11\pm 3)\times10^{-4}$. To test the sensitivity of these results to the assumed cosmological parameter distribution, we repeated the analysis using Gaussian random cosmological parameter sets based on Planck \citep{PLANCK20} with dispersion values of $\sim 0.05$. The results we obtained were IC=$(13\pm2)\times10^{-4}$. After considering all cases, we chose a value of 
\begin{equation*}
    \text{IC}=(12\pm 3)\times10^{-4}
\end{equation*}
as the most appropriate IC estimate for our study with the minitiles strategy.

\begin{figure}[h]
    \centering
    \includegraphics[width=0.5\textwidth]{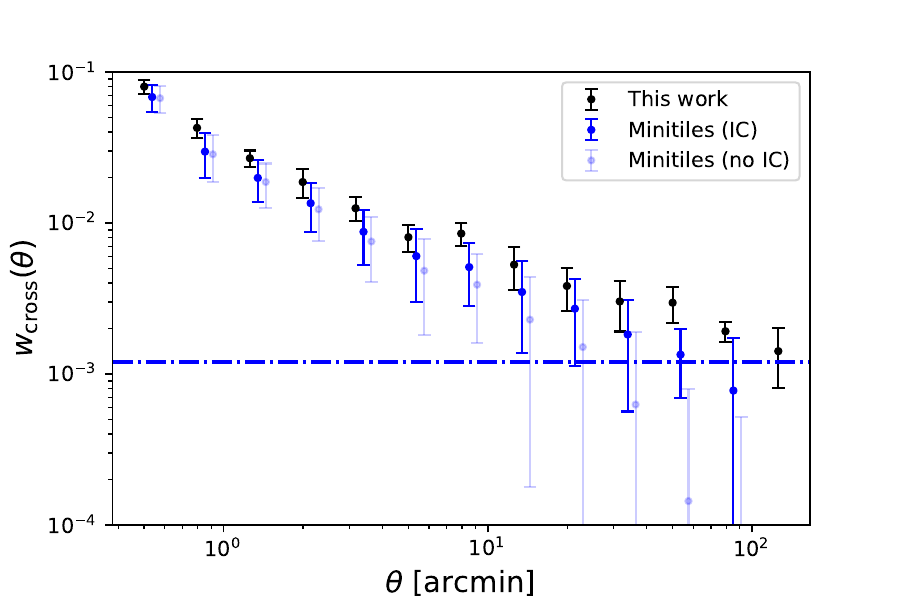}
    \caption{Angular cross-correlation data as measured in this work (in black) and using the minitiles strategy from \cite{GON21} (in blue). The light blue points represent the measurements without the IC correction, which is depicted by the dot-dashed line.}
    \label{xc_data_IC}
\end{figure}

Figure \ref{xc_data_IC} provides a comparison of the cross-correlation function estimated using different approaches. Light blue circles represent the minitiles results before the IC correction, which show a sharp decline above 10 arcmin (where the blue dashed line, representing the IC correction, becomes relevant). Blue circles correspond to the minitiles results after applying the IC correction, while the black ones are estimated using the new approach without any further correction. The uncertainties are derived from the covariance matrix, as described in the main text.

The minitiles results are consistent with the new results within the uncertainties up to 30 arcmin, but they appear to be slightly underestimated across all angular scales. Above 30 arcmin, which are the most important for cosmological constraints, the new results are clearly higher than those of the minitiles. When we consider higher IC values, such as IC=$15-20\times10^{-4}$, the two measurements become compatible even at the largest angular separations. We conclude that the IC correction procedure is not at all straightforward, since it depends on relatively arbitrary choices, such as the angular scales over which we may fit the data or the cosmologies sampled, which might be restricted but depend on a range dictated by the power-law fit. Therefore, we decided to avoid this methodology given the availability of a better alternative, which we chose for  the purposes of this work.

\section{Additional plots and tables}

\begin{figure*}[h]
    \centering
    \includegraphics[width=0.85\textwidth]{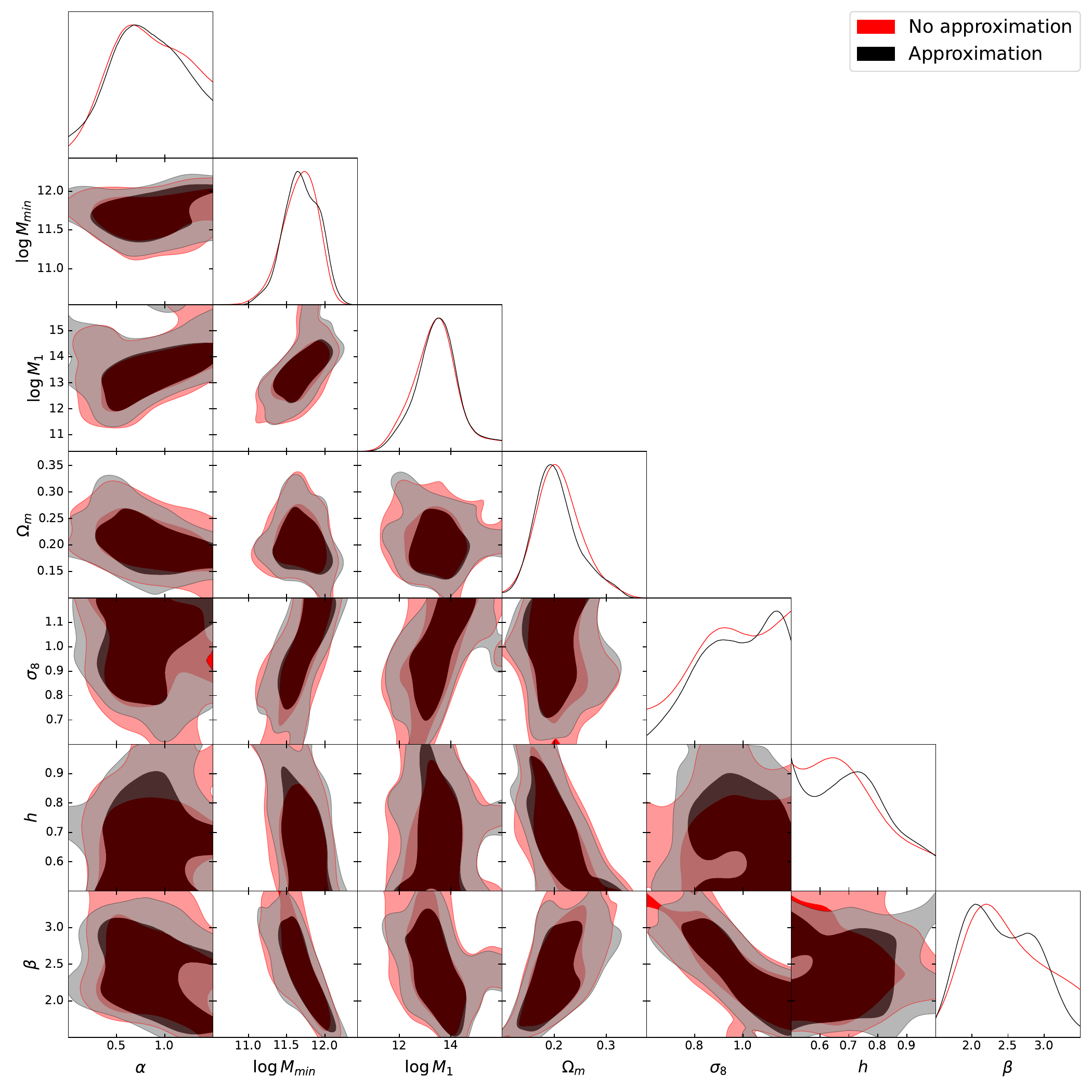}
    \caption{Marginalized posterior distributions and probability contours for the MCMC runs on the cross-correlation function with a uniform $\beta$ prior. The results using the mean-redshift approximation are shown in red, while the ones with the full model are depicted in black.}
    \label{corner_approx}
\end{figure*}

\begin{table*}[h]
\caption{Parameter prior distributions and summarized posterior results from the MCMC runs on the cross-correlation function with a uniform $\beta$ prior using the mean redshift approximation and the full model.} 
\label{table_approximation} 
\centering 
\begin{tabular}{c c c c c c c c c} 
\hline 
\hline \\[-1.2ex]
\multicolumn{1}{c}{} & \multicolumn{4}{c}{Approximation} & \multicolumn{4}{c}{No approximation} \\ 
\cmidrule(rl){2-5} \cmidrule(rl){6-9}
Parameter&Prior&Mean & Mode & 68\% CI & Prior & Mean & Mode & 68\% CI \\ 
\hline 
\\[-1ex]
$\alpha$ & $\mathcal{U}[0.00,1.50]$ & $0.83$ & $0.64$ & $[0.45,1.24]$ & $\mathcal{U}[0.00,1.50]$ &  $0.81$ & $0.71$ & $[0.56,1.50]$\\     
$\log M_{\text{min}}$ & $\mathcal{U}[10.00,16.00]$ & $11.67$ &  $11.73$ & $[11.48,11.94]$ & $\mathcal{U}[10.00,16.00]$ & $11.70$ & $11.66$ & $[11.46,11.95]$\\ [0.3ex]  
$\log M_1$ & $\mathcal{U}[10.00,16.00]$ & $13.41$ & $13.57$ & $[12.60,14.23]$ & $\mathcal{U}[10.00,16.00]$& $13.50$ & $13.56$ & $[12.68,14.29]$ \\   
$\Omega_m$ & $\mathcal{U}[0.10,1.00]$ & $0.21$ & $0.20$ & $[0.16,0.24]$ & $\mathcal{U}[0.10,1.00]$ & $0.21$ & $0.19$ & $[0.16,0.24]$ \\
$\sigma_8$ & $\mathcal{U}[0.60,1.20]$ & $0.95$ & $-$ & $[0.87,1.20]$ & $\mathcal{U}[0.60,1.20]$ & $0.97$ & $-$ & $[0.89,1.20]$ \\
$h$ & $\mathcal{U}[0.50,1.00]$ & $0.70$ & $0.65$ & $[0.50,0.75]$ & $\mathcal{U}[0.50,1.00]$ & $0.71$ & $-$ & $[0.50,0.77]$\\
$\beta$ & $\mathcal{U}[1.50,3.50]$ & $2.46$ & $2.16$ & $[1.81,2.86]$ & $\mathcal{U}[1.50,3.50]$ & $2.40$ & $2.09$ & $[1.82,2.84]$ \\
\\[-1ex]
\hline 
\hline 
\end{tabular} 
\label{table_approx}
\end{table*}

\newpage
\begin{figure*}[h]
    \centering
    \includegraphics[width=0.87\textwidth]{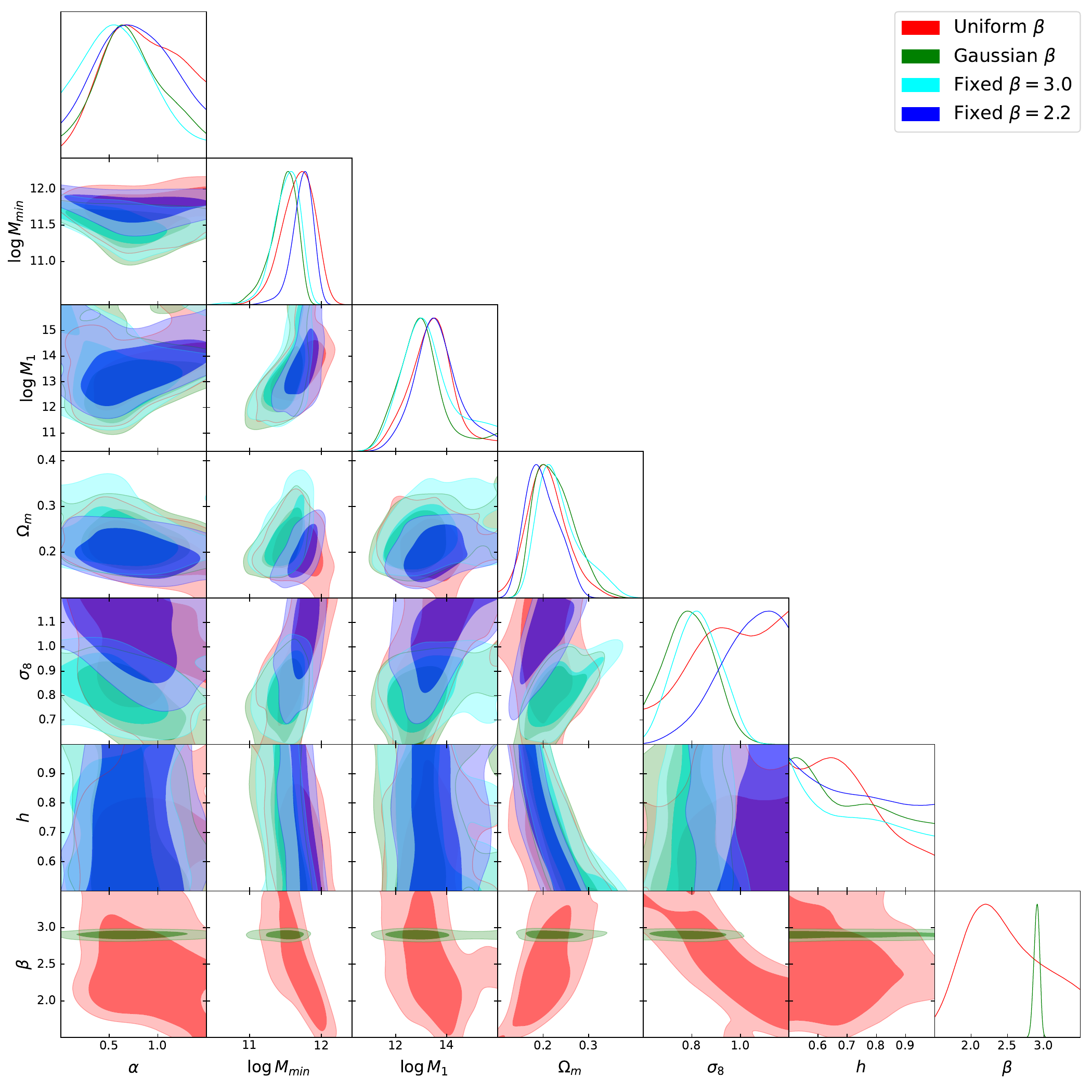}
    \caption{Marginalized posterior distributions and probability contours for the MCMC runs on the cross-correlation function for different choices of prior distributions on the $\beta$ parameter.}
    \label{corner_xc_uniform_gaussian_fixed_beta}
\end{figure*}

\begin{table*}[h]
\caption{Parameter prior distributions and summarized posterior results from the MCMC runs on the cross-correlation function with fixed values of the $\beta$ parameter: 3.0 and 2.2.} 
\centering 
\begin{tabular}{c c c c c c c c c} 
\hline 
\hline \\[-1.2ex]
\multicolumn{1}{c}{} & \multicolumn{4}{c}{Fixed $\beta=3.0$} & \multicolumn{4}{c}{Fixed $\beta=2.2$} \\ 
\cmidrule(rl){2-5} \cmidrule(rl){6-9}
Parameter&Prior&Mean & Mode & 68\% CI & Prior & Mean & Mode & 68\% CI \\ 
\hline 
\\[-1ex]
$\alpha$ & $\mathcal{U}[0.00,1.50]$ & $0.61$ & $0.54$ & $[0.18,0.90]$ & $\mathcal{U}[0.50,1.50]$ &  $0.76$ & $0.69$ & $[0.37,0.89]$\\     
$\log M_{\text{min}}$ & $\mathcal{U}[10.00,16.00]$ & $11.51$ &  $11.58$ & $[11.37,11.73]$ & $\mathcal{U}[10.00,16.00]$ & $11.74$ & $11.77$ & $[11.62,11.90]$\\      
$\log M_1$ & $\mathcal{U}[10.00,16.00]$ & $13.28$ & $13.02$ & $[12.00,13.99]$ & $\mathcal{U}[10.00,16.00]$& $13.49$ & $13.62$ & $[12.66,14.37]$ \\   
$\Omega_m$ & $\mathcal{U}[0.10,1.00]$ & $0.24$ & $0.21$ & $[0.18,0.26]$ & $\mathcal{U}[0.10,1.00]$ & $0.20$ & $0.19$ & $[0.15,0.23]$ \\
$\sigma_8$ & $\mathcal{U}[0.60,1.20]$ & $0.82$ & $0.82$ & $[0.72,0.92]$ & $\mathcal{U}[0.60,1.20]$ & $1.02$ & $1.12$ & $[0.98,1.20]$ \\
$h$ & $\mathcal{U}[0.50,1.00]$ & $0.72$ & $-$ & $[0.50,0.81]$ & $\mathcal{U}[0.50,1.00]$ & $0.73$ & $-$ & $[0.50,1.00]$\\
\\[-1ex]
\hline 
\hline 
\end{tabular} 
\label{table2}
\end{table*} 

\newpage

\begin{figure*}[h]
    \centering
    \includegraphics[width=0.85\textwidth]{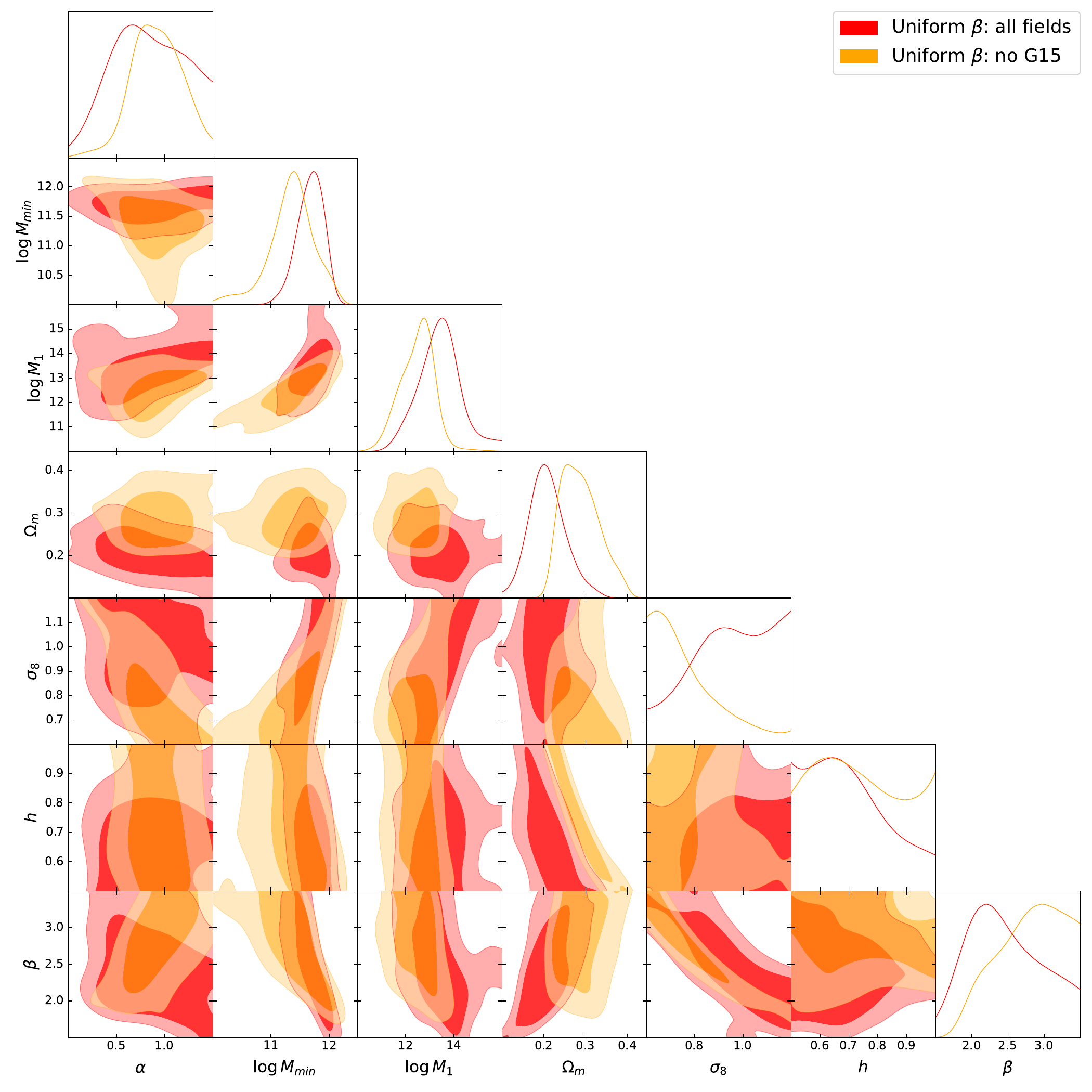}
    \caption{Marginalized posterior distributions and probability contours for the MCMC runs on the cross-correlation function with a uniform $\beta$ prior. The results using all four fields are shown in red, while those where the G15 field has been excluded are depicted in orange}.
    \label{corner_xc_uniform_all_vs_noG15}
\end{figure*}

\begin{table*}[h]
\caption{Parameter prior distributions and summarized posterior results from the MCMC run on the cross-correlation function with a uniform prior on $\beta$ and where the G15 regions was excluded from the measurement.} 

\centering 
\begin{tabular}{c c c c c} 
\hline 
\hline \\[-1.2ex]
\multicolumn{1}{c}{} 
Parameter&Prior&Mean & Mode & 68\% CI \\
\hline
\\[-1ex]
$\alpha$ & $\mathcal{U}[0.00,1.50]$ & $0.91$ & $0.86$ & $[0.65,1.19]$ \\     
$\log M_{\text{min}}$ & $\mathcal{U}[10.00,16.00]$ & $11.35$ &  $11.38$ & $[11.01,11.75]$ \\      
$\log M_1$ & $\mathcal{U}[10.00,16.00]$ & $12.49$ & $12.72$ & $[11.86,13.24]$ \\   
$\Omega_m$ & $\mathcal{U}[0.10,1.00]$ & $0.29$ & $0.27$ & $[0.23,0.32]$ \\
$\sigma_8$ & $\mathcal{U}[0.60,1.20]$ & $0.77$ & $-$ & $[0.60,0.81]$  \\
$h$ & $\mathcal{U}[0.50,1.00]$ & $0.74$ & $-$ & $[0.50,1.00]$\\
$\beta$ & $\mathcal{U}[1.50,3.50]$ & $2.78$ & $3.00$ & $[2.58,3.50]$ \\
\\[-1ex]
\hline 
\hline 
\end{tabular} 
\label{table3}
\end{table*}

\newpage

\begin{figure*}[h]
    \centering
    \includegraphics[width=0.85\textwidth]{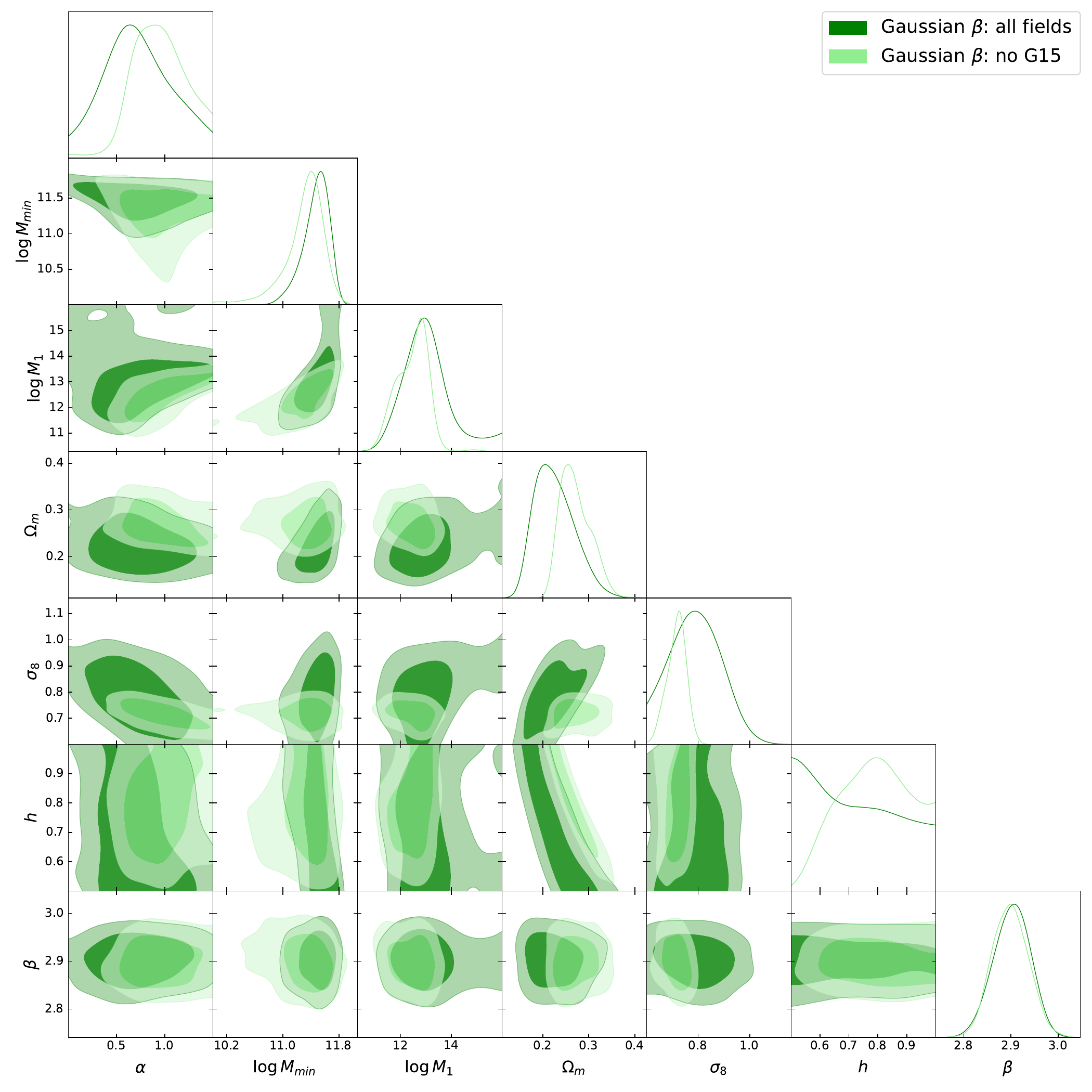}
    \caption{Marginalized posterior distributions and probability contours for the MCMC runs on the cross-correlation function with a Gaussian $\beta$ prior. The results using all fields are shown in dark green, while those where the G15 field has been excluded are depicted in light green}.
    \label{corner_xc_gaussian_all_vs_noG15}
\end{figure*}

\newpage

\begin{figure*}[h]
    \centering    \includegraphics[width=0.98\textwidth]{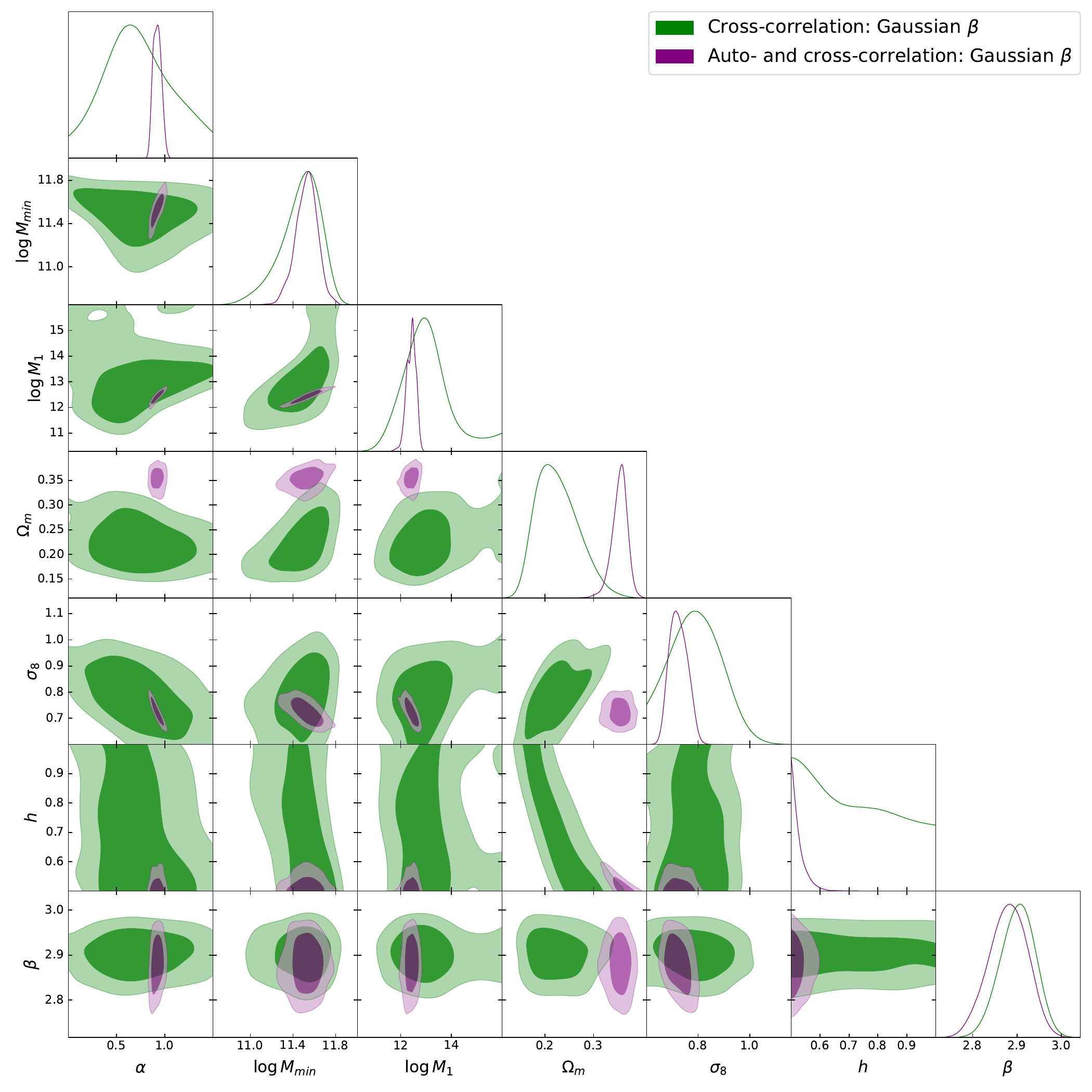}
    \caption{Marginalized posterior distributions and probability contours for the joint MCMC run on the cross- and auto-correlation functions in purple, compared with the corresponding ones using only the cross-correlation function in green. Both cases were run with a Gaussian prior on $\beta$.}
    \label{corner_xc_ac_gaussian}
\end{figure*}

\begin{table*}[h]
\caption{Parameter prior distributions and summarized posterior results from the MCMC run on the cross-correlation function with a Gaussian prior on $\beta$ and where the G15 regions was excluded from the measurement.} 

\centering 
\begin{tabular}{c c c c c} 
\hline 
\hline \\[-1.2ex]
\multicolumn{1}{c}{} 
Parameter&Prior&Mean & Mode & 68\% CI \\
\hline
\\[-1ex]
$\alpha$ & $\mathcal{U}[0.00,1.50]$ & $0.94$ & $0.91$ & $[0.63,1.19]$ \\     
$\log M_{\text{min}}$ & $\mathcal{U}[10.00,16.00]$ & $11.31$ &  $11.41$ & $[11.15,11.62]$ \\      
$\log M_1$ & $\mathcal{U}[10.00,16.00]$ & $12.46$ & $12.82$ & $[11.93,13.18]$ \\   
$\Omega_m$ & $\mathcal{U}[0.10,1.00]$ & $0.27$ & $0.26$ & $[0.23,0.29]$ \\
$\sigma_8$ & $\mathcal{U}[0.60,1.20]$ & $0.72$ & $0.73$ & $[0.68,0.76]$  \\
$h$ & $\mathcal{U}[0.50,1.00]$ & $0.79$ & $0.80$ & $[0.65,0.92]$\\
$\beta$ & $\mathcal{U}[1.50,3.50]$ & $2.90$ & $2.90$ & $[2.86,2.94]$ \\
\\[-1ex]
\hline 
\hline 
\end{tabular} 
\label{table4}
\end{table*}

\begin{table*}[h]
\caption{Parameter prior distributions and summarized posterior results from the joint MCMC run on the auto- and cross-correlation functions with a Gaussian prior on $\beta$.} 
 
\centering 
\begin{tabular}{c c c c c} 
\hline 
\hline \\[-1.2ex]
\multicolumn{1}{c}{} 
Parameter&Prior&Mean & Mode & 68\% CI \\
\hline
\\[-1ex]
$\alpha$ & $\mathcal{U}[0.00,1.50]$ & $0.92$ & $0.92$ & $[0.87,0.96]$ \\     
$\log M_{\text{min}}$ & $\mathcal{U}[10.00,16.00]$ & $11.53$ &  $11.54$ & $[11.43,11.63]$ \\      
$\log M_1$ & $\mathcal{U}[10.00,16.00]$ & $12.41$ & $12.46$ & $[12.24,12.64]$ \\   
$\Omega_m$ & $\mathcal{U}[0.10,1.00]$ & $0.36$ & $0.36$ & $[0.34,0.37]$ \\
$\sigma_8$ & $\mathcal{U}[0.60,1.20]$ & $0.72$ & $0.71$ & $[0.69,0.76]$  \\
$h$ & $\mathcal{U}[0.50,1.00]$ & $0.54$ & $-$ & $[0.50,0.55]$\\
$\beta$ & $\mathcal{N}[2.90,0.04]$ & $2.88$ & $2.88$ & $[2.84,2.93]$ \\
\\[-1ex]
\hline 
\hline 
\end{tabular} 
\label{table5}
\end{table*} 

\newpage

\begin{figure*}[h]
    \centering
    \includegraphics[width=0.92\textwidth]{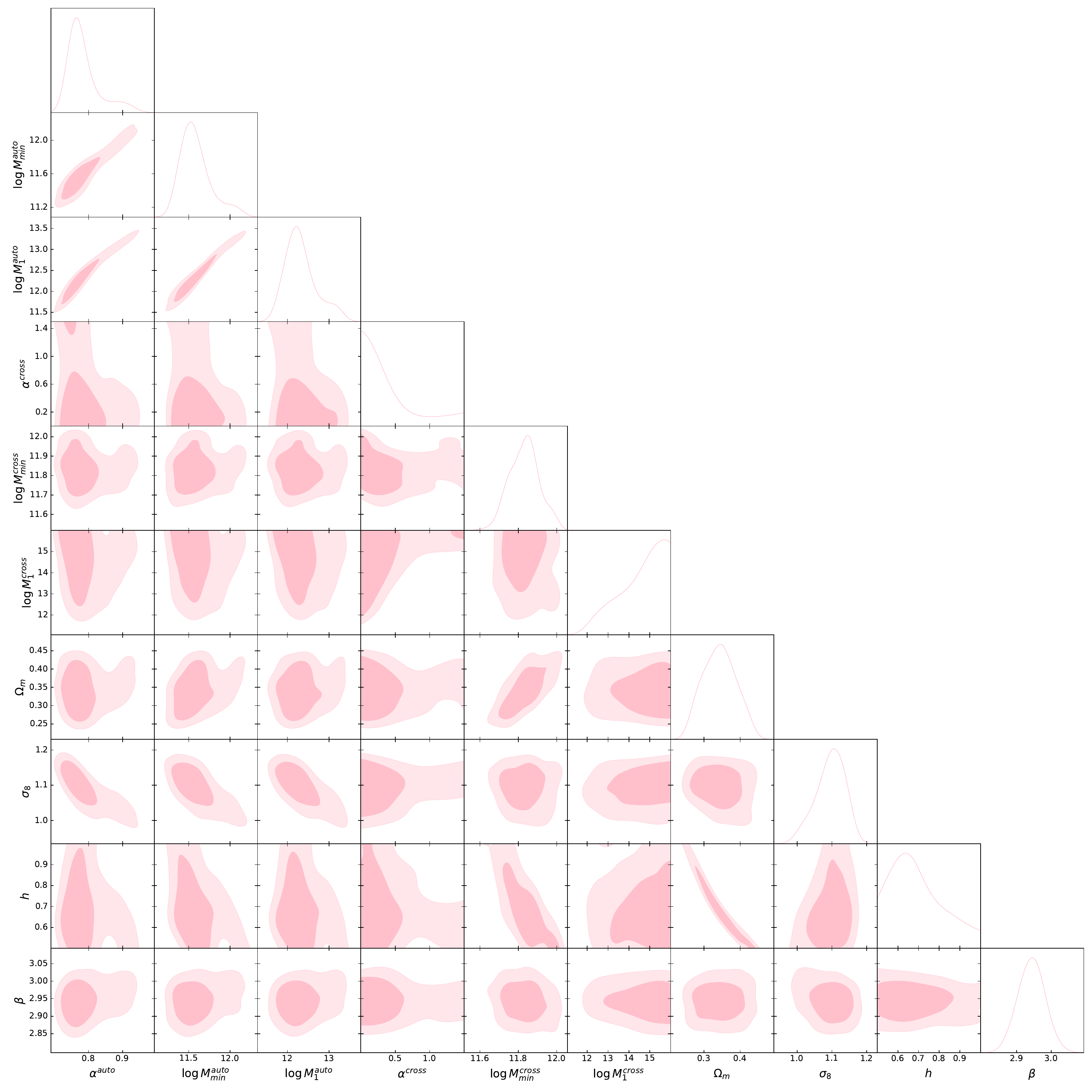}
    \caption{Marginalized posterior distributions and probability contours for the MCMC run on the cross-and auto-correlation functions assuming different HOD models for each observable.}
    \label{xc_ac_cornerplot_dHOD}
\end{figure*}

\begin{table*}[h]
\caption{Parameter prior distributions and summarized posterior results from the joint MCMC run on the auto- and cross-correlation functions assuming different HOD models for each observable.} 
 
\centering 
\begin{tabular}{c c c c c} 
\hline 
\hline \\[-1.2ex]
\multicolumn{1}{c}{} 
Parameter&Prior&Mean & Mode & 68\% CI \\
\hline
\\[-1ex]
$\alpha^{\text{auto}}$ & $\mathcal{U}[0.00,1.50]$ & $0.79$ & $0.76$ & $[0.73,0.80]$ \\     
$\log M^{\text{auto}}_{\text{min}}$ & $\mathcal{U}[10.00,16.00]$ & $11.58$ &  $11.52$ & $[11.36,11.70]$ \\      
$\log M^{\text{auto}}_1$ & $\mathcal{U}[10.00,16.00]$ & $12.33$ & $12.22$ & $[11.86,12.57]$ \\   

$\alpha^{\text{cross}}$ & $\mathcal{U}[0.00,1.50]$ & $0.41$ & $-$ & $[0.00,0.43]$ \\     
$\log M^{\text{cross}}_{\text{min}}$ & $\mathcal{U}[10.00,16.00]$ & $11.83$ &  $11.85$ & $[11.75,11.91]$ \\      
$\log M^{\text{cross}}_1$ & $\mathcal{U}[10.00,16.00]$ & $14.50$ & $-$ & $[14.08,16.00]$ \\   

$\Omega_m$ & $\mathcal{U}[0.10,1.00]$ & $0.34$ & $0.35$ & $[0.29,0.39]$ \\
$\sigma_8$ & $\mathcal{U}[0.60,1.20]$ & $1.10$ & $1.11$ & $[1.06,1.15]$  \\

$h$ & $\mathcal{U}[0.50,1.00]$ & $0.69$ & $0.64$ & $[0.52,0.75]$\\
$\beta$ & $\mathcal{N}[2.90,0.04]$ & $2.94$ & $2.95$ & $[2.90,2.98]$ \\
\\[-1ex]
\hline 
\hline 
\end{tabular} 
\label{table6}
\end{table*} 

\begin{figure*}[h]
\centering
    \includegraphics[width=0.5\textwidth]{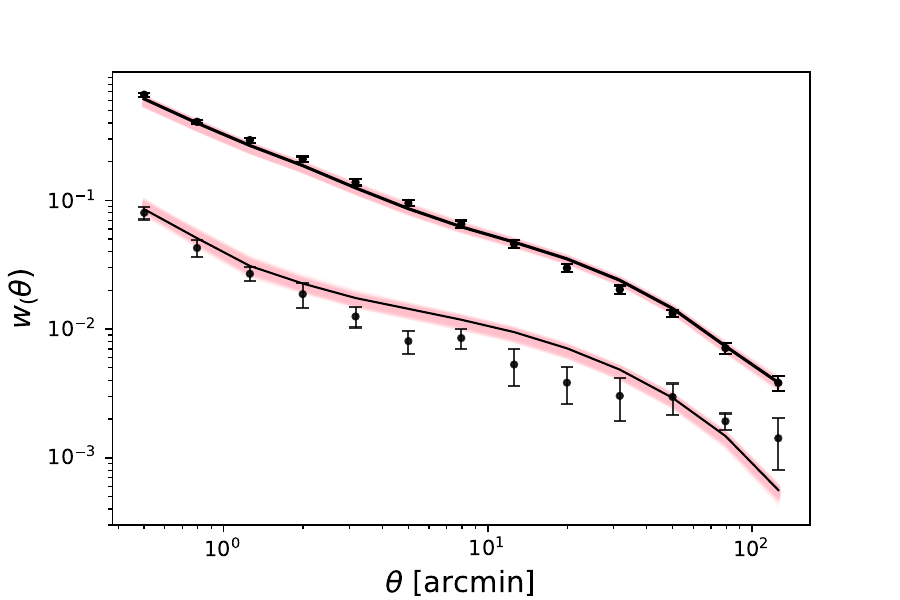}
    \caption{Posterior-sampled angular auto- and cross-correlation functions from the joint analysis with different HOD parameters. The data are shown in black.}
    \label{xc_ac_sampling_dHOD}
    
\end{figure*}

\end{document}